\documentclass[fleqn,usenatbib]{mnras}

\usepackage{newtxtext,newtxmath}
\usepackage[T1]{fontenc}

\DeclareRobustCommand{\VAN}[3]{#2}
\let\VANthebibliography\thebibliography
\def\thebibliography{\DeclareRobustCommand{\VAN}[3]{##3}\VANthebibliography}

\usepackage{graphicx}	% Including figure files
\usepackage{amsmath}	% Advanced maths commands

\usepackage{multirow}
\usepackage{braket}
\usepackage{comment}
\usepackage{bm}
\usepackage{soul}
\usepackage{xcolor}

\title[Vanishing EoR cross power]{Insights into the 21 cm field from the vanishing cross-power spectrum at the epoch of reionization}

\author[K. Moriwaki et al.]{
Kana Moriwaki,$^{1,2}$\thanks{E-mail: kana.moriwaki@phys.s.u-tokyo.ac.jp}
Angus Beane,$^{3}$
and Adam Lidz$^{4}$
\\
$^{1}$Research Center for the Early Universe, Graduate School of Science, The University of Tokyo, 7-3-1 Hongo, Bunkyo, Tokyo 113-0033, Japan\\
$^{2}$Department of Physics, Graduate School of Science, The University of Tokyo, 7-3-1 Hongo, Bunkyo, Tokyo 133-0033, Japan\\
$^{3}$Center for Astrophysics $|$ Harvard \& Smithsonian,  Cambridge, MA 02138, USA\\
$^{4}$Department of Physics and Astronomy, University of Pennsylvania, 209 South 33rd Street, Philadelphia, PA 19104, USA\\
}

\date{Accepted XXX. Received YYY; in original form ZZZ}

\pubyear{2024}

\begin{document}
\label{firstpage}
\pagerange{\pageref{firstpage}--\pageref{lastpage}}
\maketitle

\begin{abstract}
The early stages of the Epoch of Reionization, probed by the 21 cm line, are sensitive to the detailed properties and formation histories of the first galaxies. 
We use {\sc 21cmFAST} and a simple, self-consistent galaxy model to examine the redshift evolution of the large-scale cross-power spectrum between the 21 cm field and line-emitting galaxies. 
A key transition in redshift occurs when the 21 cm field shifts from being positively correlated with the galaxy distribution to being negatively correlated. Importantly, this transition redshift is insensitive to the properties of the galaxy tracers but depends sensitively on the thermal and ionization histories traced through the 21 cm field.
Specifically, we show that the transition occurs when {\em both} ionization fluctuations dominate over 21 cm spin temperature fluctuations and when the average spin temperature exceeds the temperature of the cosmic microwave background. 
We illustrate this with three different 21 cm models which have largely the same neutral fraction evolution but different heating histories.
We find that the transition redshift has a scale dependence, and that this can help disentangle the relative importance of heating and ionization fluctuations. 
The best prospects for constraining the transition redshift occur in scenarios with late X-ray heating, where the transition occurs at redshifts as low as $z \sim 6-8$.
In our models, this requires high-redshift galaxy surveys with sensitivities of 
$\sim 10^{-18}~\rm erg/s/cm^2$ for optical lines and $\sim 10^{-19}~\rm erg/s/cm^2$ for FIR lines. 
Future measurements of the transition redshift can help discriminate between 21 cm models and will benefit from reduced systematics.
\end{abstract}


%%%%%%%%%%%%%%%%%%%%%%%%%%%%%%%%%%%%%%%%%%%%%%%%%%

%%%%%%%%%%%%%%%%% BODY OF PAPER %%%%%%%%%%%%%%%%%%

\section{Introduction}
The redshifted 21 cm line contains a great deal of information about the structure and evolution of the Epoch of Reionization (EoR). 
Measurements of the EoR at $z \sim 6-15$ will reveal the nature of the first galaxies and black holes, as well as the properties of large-scale structure at high redshift \citep{Loeb13}.
While instruments are nearing the sensitivity necessary to detect the EoR signal, foreground contamination and instrumental artifacts have so far resulted in only upper limits on the amplitude of 21 cm fluctuations 
\citep[e.g.,][]{Paciga13, %GMRT
Dillon14, Beardsley16, Ewall-Wice16, Barry19, Trott20, Yoshiura21, %MWA
Patil17, Gehlot19, Gehlot20, Mertens20, %LOFAR
Eastwood19, %LWA
HERA22}. 

Because of the great difficulties confronting 21 cm auto-spectrum measurements, researchers are exploring the prospects for measuring 21 cm signals in cross-correlation with another tracer of large-scale structure, such as galaxies. 
The 21 cm-galaxy cross-correlations were first explored by \citet{Wyithe07} in the context of using Ly$\rm \alpha$-emitters to distinguish between inside-out and outside-in reionization scenarios. 
A number of authors have studied the detectability of the large scale anti-correlation expected in cross-correlation during the EoR \citep[e.g.,][]{Vrbanec16, Hutter17, Kubota18, Weinberger20, Padmanabhan23,LaPlante23}, and its dependence on reionization models \citep[e.g.,][]{Wiersma13,Park14,Hutter23}. 
Most studies have been done for Ly$\rm \alpha$-emitters, for which the observational range is limited to the late stages of reionization since they are obscured by neutral hydrogen \citep[e.g.,][]{Kashikawa06}.
However, there are several studies estimating the cross-correlation signals with alternative tracers such as line emitters or line intensity maps of CO \citep{Lidz11}, [CII] \citep{Gong12}, Ly$\rm \alpha$ \citep{Silva13}, H$\rm \alpha$ \citep{Neben17,Heneka21}, and [OIII] \citep{Moriwaki19}. 
Using such emission lines may allow us to probe higher redshifts.
The possibility of combining 21 cm and two additional emission lines has also been explored \citep{Beane19} along with using the cross-bispectrum between 21 cm and another density tracer \citep{Beane18}.

The redshift evolution of the cross-power spectrum and its dependence on the reioniziation model has also been predicted and studied \citep[e.g.,][]{Lidz09,Gong12,Wiersma13,Park14,Vrbanec16,Dumitru19,Kannan22}.
However, in the realms of the other 21 cm statistics, such as global 21 cm signal \citep[e.g.,][]{Cohen17}, 21 cm auto-power spectrum \citep[e.g.,][]{Fialkov14,Cohen18,Park19}, and other higher-order statistics \citep[e.g.,][]{Wyithe07,Shimabukuro17,Kamran21}, a significant approach has been the tracking of their continuous evolution, i.e., charting the signals as a function of redshift,
which has not yet been extensively explored in cross-power spectrum studies \citep[although see][for cross-correlation function]{Heneka20}. Such studies have proven vital in identifying at which redshift significant characteristics appear in the signals and in elucidating the relationship with reionization models. Ultimately, this approach will be invaluable for comparing theoretical predictions with observational data. 
Here we aim to quantify the analogous continuous evolution of 21 cm cross-power spectrum signals.

In general, the detection of small-scale (large-$k$) cross-power spectra is more challenging because of the limited angular resolution of 21 cm survey and redshift uncertainties associated with galaxy surveys \citep[e.g.,][]{Kubota18,Moriwaki19}.
The largest wavenumbers (finest resolution) probable with upcoming observations at the EoR
will be approximately $k_{\perp, \rm max} \sim k_{||, \rm max} \sim 1~h~\rm Mpc^{-1}$ assuming an SKA-like 21 cm observation and a galaxy survey with redshift uncertainty of $\sigma_z \sim 0.01$.\footnote{For an interferometer, the longest baseline $b_{\rm max}$ determines the resolution as $k_{\perp, \rm max} = 2\pi b_{\rm max}/ \lambda_{21} / \chi$, where $\lambda_{21} = 21~\rm cm$, and $\chi$ is the comoving distance. Assuming $b_{\rm max} \sim 1$ km for SKA, we obtain $k_{\perp, \rm max} \sim 2~h~\rm Mpc^{-1}$. With the 21 cm observation alone, smaller scales can still be explored thanks to the very high spectral resolution of the interferometer.
In a galaxy survey, however, the redshift uncertainty relates to the maximum line-of-sight wavenumber as $k_{||, \rm max} = H(z) /c \sigma_z$, where $c$ is the speed of light. The uncertainty of $\sigma_z = 0.01$ corresponds to $k_{||, \rm max} \sim 0.7~h~\rm Mpc^{-1}$ at $z = 10$.}
Another challenge with small-scale signals is their dependence on modeling  small-scale physics (e.g., clumping factor), which are challenging to resolve
in current simulations.
Therefore, in this paper, we focus on the large-scale cross-power spectrum and study its general behaviour for different 21 cm models. 

Cross-power spectrum prediction necessitates modeling the populations of the density tracers (galaxies) as well as the 21 cm field. 
In this study, we adopt a very simple, yet reasonable and realistic galaxy model that allows us to discuss separately the general trend of the cross-power spectrum and its dependence on the properties of individual emission lines.
In Sec.\ref{sec:method}, we describe our methodology for generating the 21 cm and galaxy fields. 
For simplicity, we only consider three different 21 cm models which have largely the same neutral fraction evolution but different heating histories.
Sec.\ref{sec:result} outlines the general evolutionary trends of the cross-power spectra on large scales. We initiate our discussion by characterizing the galaxy population in terms of halo mass, a parameter readily obtained from the simulation.
Sec.\ref{sec:discussion} then discusses the signal detectability and the required line sensitivities. We consider four emission lines, H$\rm \alpha$, [OIII]5007\AA, [OIII]88$\rm \mu$m, and [CII]158$\rm \mu$m.
Throughout this paper, we adopt a $\rm \Lambda$ cold dark matter cosmology with $\Omega_{\rm m} = 0.316$, $\Omega_{\rm b} = 0.0489$, $h = 0.673$, $\sigma_8 = 0.812$, and $n_s = 0.966$ \citep{Planck18}.

\section{Methods}
\label{sec:method}

\subsection{21 cm signals}

We generate realizations of the 21 cm field using the publicly available code {\sc 21cmFAST} v3.2 \citep{Mesinger11,Murray20}. 
The boxsize and the number of grid cells are, respectively, set to 350 co-moving $h^{-1}$ Mpc and 256$^3$. We obtain outputs at several redshifts instead of generating a light cone. The light-cone effect is known to be modest \citep{Datta12,Murmu21}.

We adopt the parametrization used by \citet{Park19}. 
The stellar masses of a galaxy in a halo with a mass of $M_{\rm h}$ are computed as 
\begin{equation}
    M_*(M_{\rm h}) = f_{\rm *,10}
    \Big(\frac{M_{\rm h}}{10^{10}M_\odot}\Big)^{\alpha_*}
    \frac{\Omega_{\rm b}}{\Omega_{\rm m}} M_{\rm h}, 
\end{equation}
where $f_{\rm *,10}$ is stellar-mass fraction in halos with $M_{\rm h} = 10^{10}~\rm M_\odot$, and $\alpha_*$ is power-law index.
The star formation rate (SFR) is then computed as 
\begin{equation} \label{eq:sfr_21cmfast}
    {\rm SFR} = \frac{M_*}{t_*H(z)^{-1}},
\end{equation}
with a dimensionless parameter for the star-formation time scale, $t_* \in [0,1]$, and Hubble constant $H(z)$.
The escape fraction of ionizing photons is allowed to vary depending on the halo mass as
\begin{equation}
    f_{\rm esc}(M_{\rm h}) = f_{\rm esc,10}\Big(\frac{M_{\rm h}}{10^{10}M_\odot}\Big)^{\alpha_{\rm esc}},
\end{equation}
with normalization coefficients $f_{\rm esc,10}$ and power-law index $\alpha_{\rm esc}$.
Star formation in small mass haloes is expected to be quenched by, e.g., stellar feedback. This effect is incorporated as a duty cycle,
\begin{equation}
    f_{\rm duty} (M_{\rm h}) = \exp \Big(-\frac{M_{\rm turn}}{M_{\rm h}}\Big).
\end{equation}
This leads to an exponential suppression in the number of small mass halos that form stars at $M_{\rm h} \ll M_{\rm turn}$, and so such halos contribute negligibly to reionization in our models. 

We adopt a set of parameters from the range allowed by current observations \citep{HERA22b}: $\log f_{\rm *,10} = -1.4$, $\alpha_* = 0.5$, $t_* = 0.4$, $\log f_{\rm esc} = -1.30$, $\alpha_{\rm esc} = 0.3$, and $\log M_{\rm turn} = 8.0$.
We confirm that the resulting galaxy populations are consistent with the observed UV luminosity functions at $z = 6- 10$ \citep{Finkelstein15, Bouwens15, Livermore17, Atek18, Bhatawdekar19}
and that the Thomson scattering optical depth to the CMB aligns closely with \citet{Planck18}.

We are interested in spin temperature ($\sim$ gas temperature at the redshifts of interest) evolution. 
While recent observations prefer somewhat early heating models \citep[e.g.,][]{HERA22b}, a wide range of models are still viable.
The X-ray emissivity is parametrized by the specific X-ray luminosity below 2 keV per unit star formation ($L_{\rm X}/{\rm SFR}$), 
the energy threshold below which X-ray photons are absorbed by the host galaxies ($E_0$), and the X-ray spectral energy index ($\alpha_{\rm X}$) \citep{Greig18}.
We adopt $E_0 = 300$ keV and $\alpha_{\rm X} = 1.5$ and 
consider three different values for X-ray luminosity, $\log \frac{L_{\rm X}/{\rm (erg/s)}}{\rm SFR/\rm (M_\odot /yr)} = 39, 40$, and 41.

\begin{figure*}
\includegraphics[width=16cm]{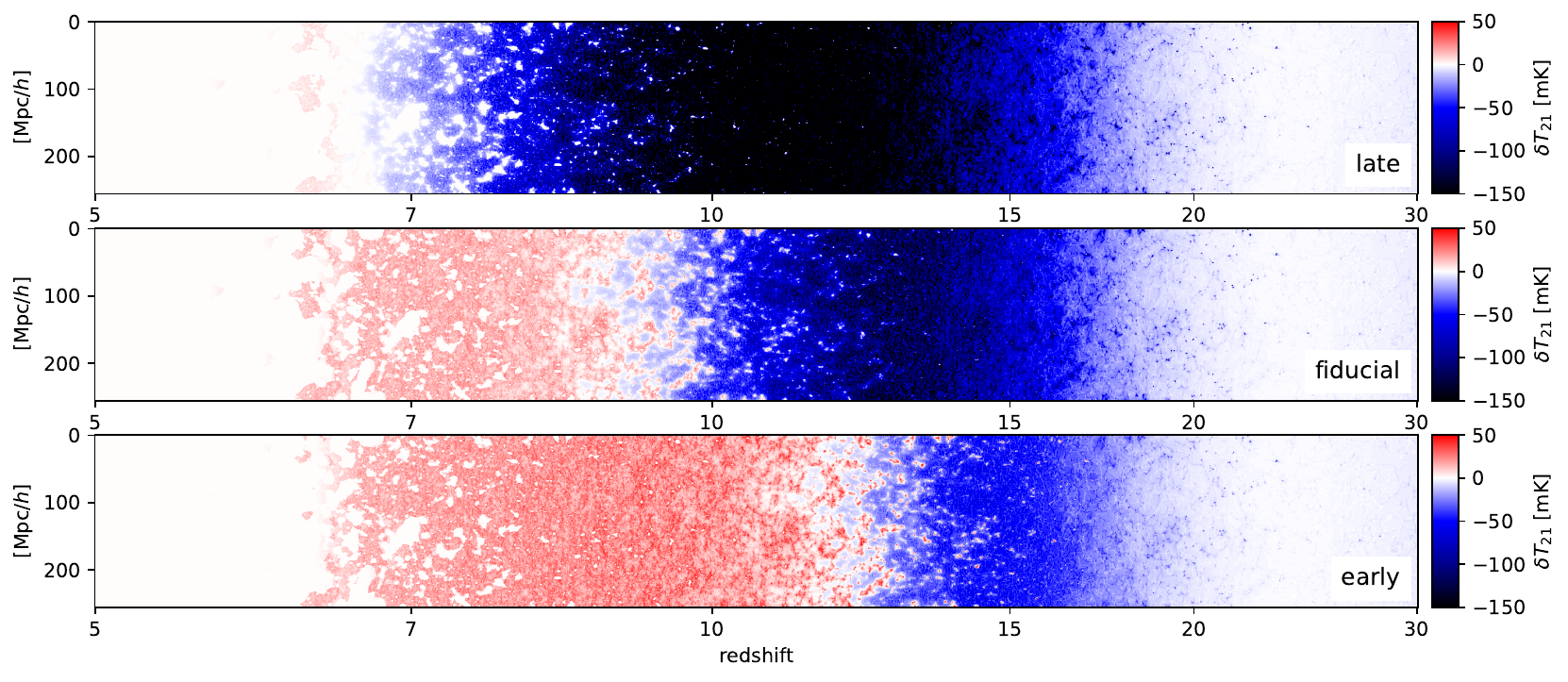}
\caption{Light cone slices showing the 21 cm brightness temperature (Eq.\ref{eq:Tb}) for three different models, each with $\log L_{\rm X}/\rm SFR = 39$ (late), 40 (fiducial), and 41 (early).}
\label{fig:lightcone}
\end{figure*}

From the simulated distributions of ionized fraction and gas temperature, 
the brightness temperature of the 21cm signal is computed as
\begin{equation} \label{eq:Tb}
	\delta T_{21} = T_0 x_{\rm HI} (1+\delta_\rho) \Big(1-\frac{T_{\rm CMB}}{T_{\rm s}}\Big) \frac{H(z)}{H(z)+\frac{dv_{||}}{dr_{||}}},
\end{equation}
where $\delta_\rho$ is the fluctuation of the gas density, 
$x_{\rm HI}$ is the neutral fraction,
$T_{\rm s}$ is the spin temperature,
$dv_{||}/dr_{||}$ is the velocity gradient along the line of sight,
and 
\begin{equation}
	T_0 = 27 \Big(\frac{\Omega_{\rm b}h^2}{0.022} \Big) \Big(\frac{1+z}{10} \frac{0.15}{\Omega_{\rm m}h^2}\Big)^{1/2}~\rm mK
\end{equation}
is a normalization factor.

Fig.\ref{fig:lightcone} shows 21 cm brightness temperature light cones for models with $\log L_{\rm X}/\rm SFR = 39$ (top), 40 (middle), and 41 (bottom), which are obtained by re-running the simulations in light cone mode. As expected, the stronger the X-ray emissivity is, the earlier the heating occurs.
We refer to the scenarios as the late heating, fiducial, and early heating models, respectively.

\subsection{Galaxies}

When running {\sc 21cmFAST}, we also generate halo fields from the same initial conditions using the halo finder included in {\sc 21cmFAST}. 
Specifically, the halos are identified using the excursion-set formalism by filtering the linear density field on a range of smoothing scales \citep{Mesinger07}. The locations of the haloes are then adjusted at each redshift using the displacement field.
The resulting halo catalog contains halos with $M_{\rm h} \gtrsim 10^{10}~\rm M_\odot$. 
Such galaxies have ${\rm SFR} \gtrsim 0.2~{\rm M_\odot}/\rm yr$ and absolute UV magnitudes of $M_{\rm UV, abs} \lesssim -17$ at $z \sim 10$. As a side note, this roughly corresponds to the current upper bounds on the JWST Phot-$z$ and Spec-$z$ samples \citep[e.g.,][]{Perez-Gonzalez23, Harikane23}.
We note that for the purposes of computing 21 cm fluctuations, {\sc 21cmFAST} models the halo collapse fraction and so ionizing photons from halos down to $M_{\rm h} \gtrsim M_{\rm turn} = 10^{8}~\rm M_\odot$ are properly accounted for.
From the obtained halo catalog, we generate number density maps $n_{\rm gal}$ for galaxies with flux above some threshold sensitivity. 

Detecting emission lines from galaxies is crucial to precisely measure their positions along the line of sight rather than relying on photometric redshifts alone \citep{Furlanetto07}. 
The luminosities of emission lines such as $\rm H\alpha$, [OIII]5007\AA, [OIII]88$\rm \mu$m, and [CII]158$\rm \mu$m correlate with the SFR, and the relation can be written as \citep[e.g.,][]{Moriwaki18}
\begin{equation} \label{eq:L-SFR}
    L_{\rm line} = C_{\rm line} (1-f_{\rm esc}) {\rm SFR}
\end{equation}
with a coefficient $C_{\rm line}$.

For simplicity, we do not specify the line considered in the majority of this paper. Instead of using a flux limit, we generally work with a halo-mass limit and discuss the mass-flux relation only in Sec.\ref{sec:flux_limit}.
This consideration is valid (or at least consistent with the reionization simulation) when the coefficient $C_{\rm line}$ is constant.
In reality, the coefficient in Eq.\ref{eq:L-SFR} depends on the properties of the interstellar medium (ISM) in each galaxy, such as the typical metallicity, density, and ionization parameter.
In practice, these will add scatter to the correlation of Eq.\ref{eq:L-SFR},
as will be investigated in Sec.\ref{sec:galpop}.
We note that the stochastic star-formation activities should also introduce scatter in the halo mass-SFR relation,
but we postpone the consideration of this effect to future studies, as it necessitates a new implementation within the simulation itself, not just in the post-processing.

\section{Results}
\label{sec:result}

From the obtained 21cm field (Eq.\ref{eq:Tb}) and 
the number density fluctuation $\delta_{\rm gal} = n_{\rm gal} / \braket{n_{\rm gal}} -1$, 
we compute the cross-power spectrum. 
We will use the normalized power spectrum
\begin{equation}
    \Delta^2(k) \equiv \frac{k^3}{2\pi^2}P(k).
\end{equation}
and a logarithmic width of ${\rm d}\log k = 0.3$ in the following.

\begin{comment}
\subsection{Bubble radius}

It has been said in several studies that the cross-correlation can be used to investigate the bubble size distribution. We check it in this section.

Fig.\ref{fig:ccf} shows mean neutral fraction (top), temperature coefficient (middle), and brightness temperature of 21cm (bottom) around galaxies with $\log L > 40.5~\rm erg/s$.
We define the bubble radius $r_{\rm bubble}$ as
\begin{equation}
    x_{\rm HI}(r_{\rm bubble}) = 0.5 * x_{\rm HI}(r = \infty),
\end{equation}
They are indicated as the vertical lines in Fig.\ref{fig:ccf}.

In Fig.\ref{fig:cps}, we show the cross-power spectra (top/middle) and correlation coefficients. 
The vertical lines shows the scale of the bubble radius $k = 2\pi / r_{\rm bubble}$.
they roughly corresponds to the scales where the cross-power spectra get zero.
This is just because there is almost no gradient of the brightness temperature within the bubble radius (see bottom panel of Fig.\ref{fig:ccf}).
Signals at these scales are actually very difficult to detect (see discussion in Section \ref{sec:discussion}). 

\end{comment}

\subsection{General behaviours of the large-scale signals}

\begin{figure}
	\includegraphics[width=8cm]{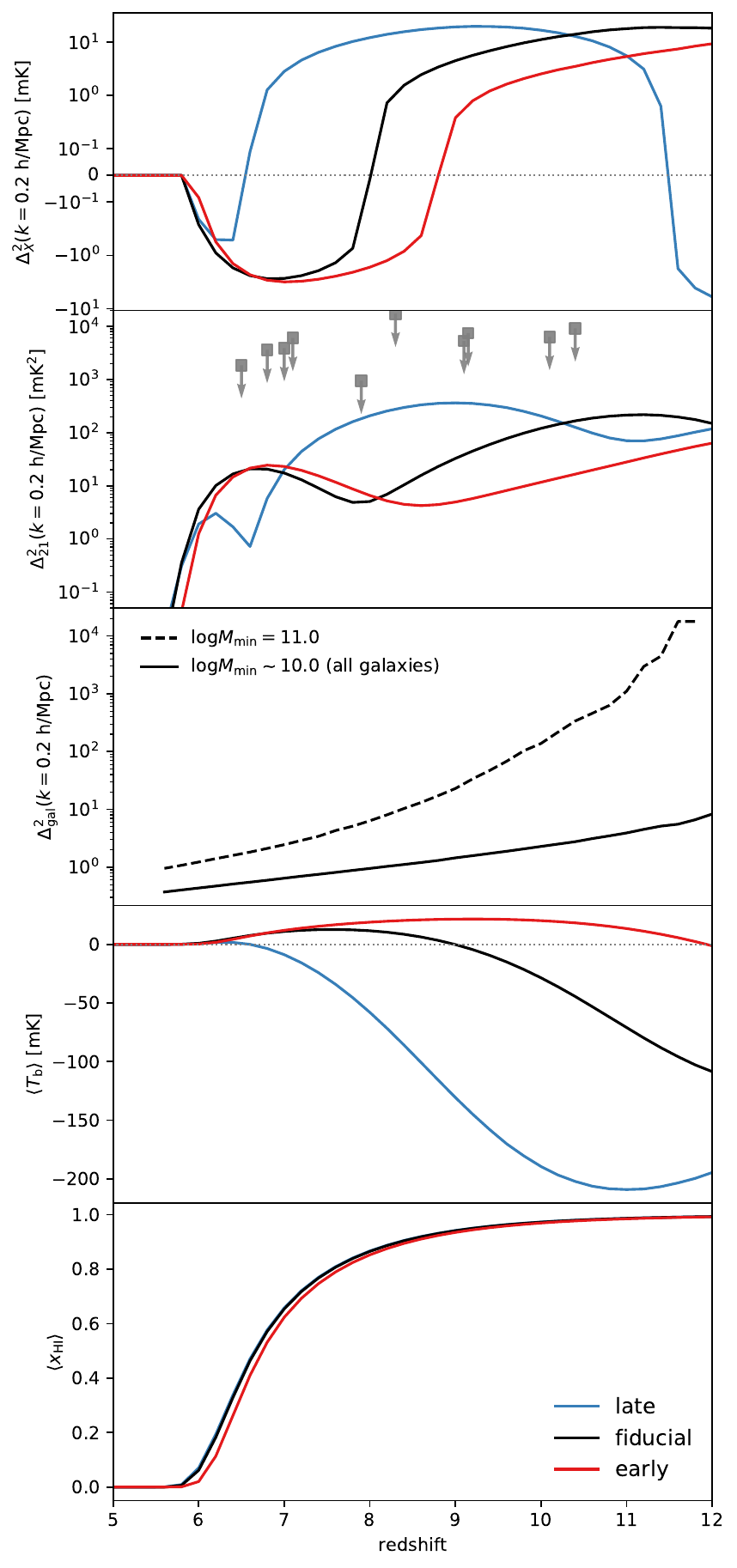}
    \caption{
    The redshift evolution of the large-scale cross-power spectra between 21 cm and all of the halos in the simulation (top), 21cm auto-power spectra (second row), galaxy auto-power spectra (third row), 21cm global signals (fourth row), and mean neutral fraction (bottom).}
    The power spectra at $k = 0.2~h~\rm Mpc^{-1}$ are shown. 
    The gray arrows in the second row show upper limits obtained in recent observations \citep{Barry19,Trott20,Mertens20,Patil17,HERA22}. 
    \label{fig:ps_zev}
\end{figure}

\begin{figure*}
    \includegraphics[width=18cm]{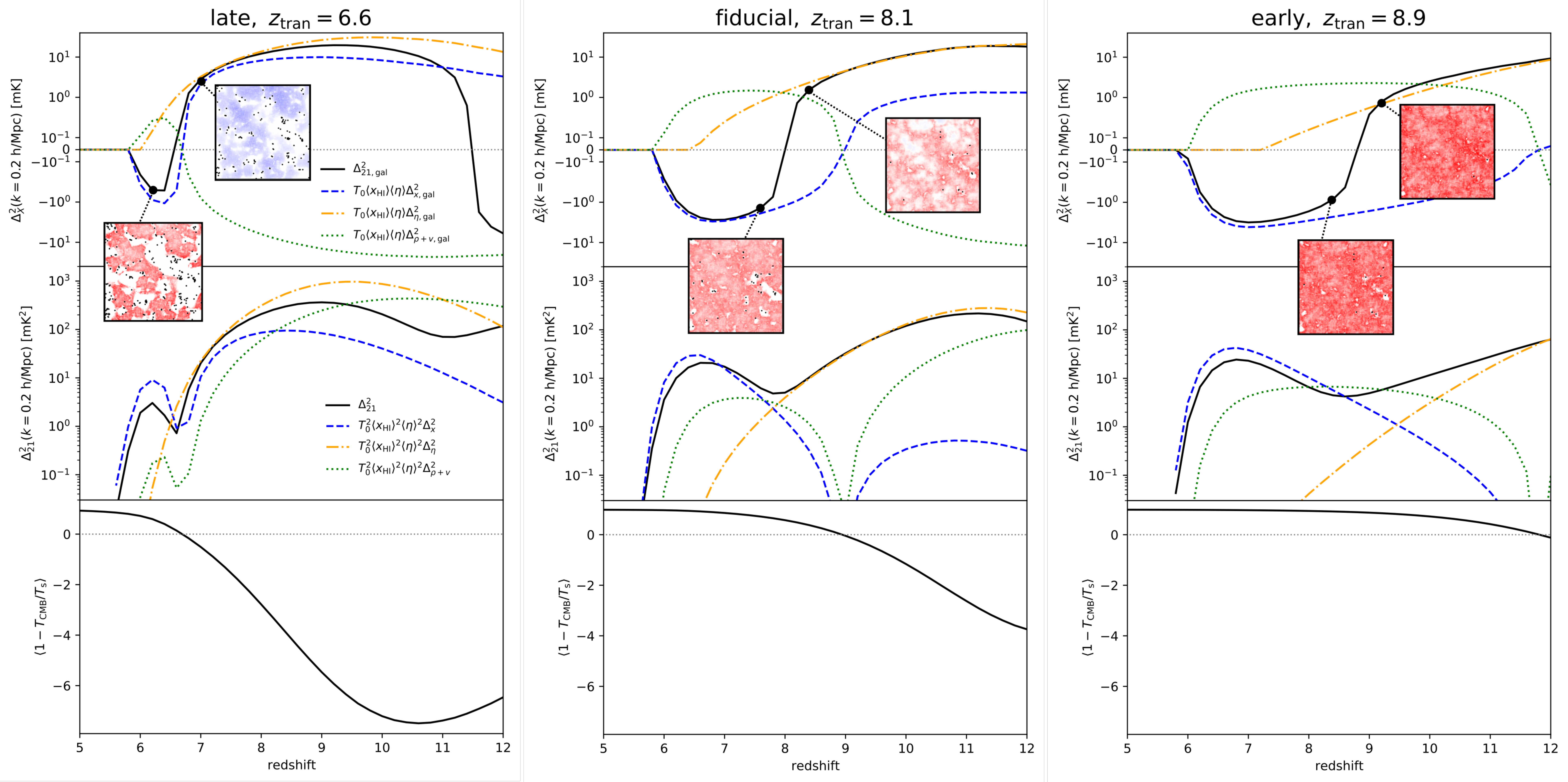}
\caption{
The cross-power spectra between 21 cm and all halos in the simulation (top) and 21cm auto-power spectra (middle row) for late heating (left), fiducial (middle), and early heating (right) models. The blue dashed, orange dash-dotted, and green dotted lines indicate the contributions from ionized, temperature, density fluctuation terms in Eq.\ref{eq:Tb_comp}. The inset color maps show slices of the 21 cm signals (red: emission, blue: absorption) and the distributions of the massive haloes (black dots) around the positive-to-negative transition redshifts denoted by $z_{\rm tran}$. 
The bottom panels show the evolution of $\braket{\eta} = \braket{1-T_{\rm CMB}/T_{\rm s}}$. The point where $\braket{\eta}$ crosses zero is known as the heating transition. 
}
\label{fig:ps_zev_comp}
\end{figure*}

The cross-power spectrum between the 21 cm and density field as a function of redshift has been investigated in, e.g., \citet{Fialkov14}. 
The cross-power spectrum between the 21 cm and biased tracers, such as galaxies, should exhibit characteristics that are similar yet distinct. 

Here we consider all the halos within the simulation to study the general behaviour of the large-scale cross-power spectrum. 
The minimum halo mass is $M_{\rm h}\sim 10^{10}~\rm M_\odot$. 
The top panel of Fig.\ref{fig:ps_zev} shows the redshift evolution of the cross-power spectra at $k = 0.2 ~h~\rm Mpc^{-1}$ for three different reionization models. 
We also show the redshift evolution of the: 21 cm auto-power spectra (second row), galaxy auto-power spectra (third row), 21 cm global signals (third row), and mean neutral fraction (bottom row).
For the galaxy power spectra, the case with a halo mass threshold of $\log M_{\rm min} = 11.0$ is included for reference.
We confirm that there is no significant difference in the reionization histories and that the observed variations arise solely from the different heating scenarios.
As a general trend, the cross-power spectra are positive at high redshifts and negative at low redshifts. 
When comparing different models, a higher X-ray emissivity (earlier heating) leads to an earlier transition from positive to negative correlation.

In order to further understand these results, we decompose the cross-power spectrum into a number of constituent terms.
Similar decompositions have been explored in the context of the auto-power spectrum \citep{Lidz07,Georgiev22} and cross-power spectrum \citep{Lidz09}.
We rewrite Eq.\ref{eq:Tb} as 
\begin{equation} \label{eq:Tb_comp}
	\delta T_{21} = T_0\, \braket{x_{\rm HI}}\, \braket{\eta}\, (1+\delta_{\rho+v})(1+\delta_x) (1+\delta_\eta) ,
\end{equation}	
where $\braket{x_{\rm HI}}$ and $\braket{\eta}$ are the volume averaged values of the neutral fraction and the temperature term, $\eta \equiv 1-T_{\rm CMB}/T_{\rm s}$.
Their fluctuations are defined as $\delta_x = x_{\rm HI}/ \braket{x_{\rm HI}} - 1$, and $\delta_\eta = \eta /\braket{\eta} -1 $, respectively, 
and $\delta_{\rho+v}$ is the density perturbation in redshift space. 
The cross-power spectrum between 21 cm and galaxies is then given as
\begin{equation}\label{eq:Px_comp}
	\Delta_{\rm 21,gal}^2 = T_0\, \braket{x_{\rm HI}}\, \braket{\eta}\, 
					(\Delta^2_{x,\rm gal} + \Delta^2_{\eta, \rm gal} + \Delta^2_{\rho+v,\rm gal}+ \cdots ). 
\end{equation}	
We omit the correlations between galaxies $\delta_{\rm gal}$ and higher order fluctuation terms such as $\delta_{\rho+v} \delta_x$.
Similarly, the auto-power spectrum is decomposed as 
\begin{equation}\label{eq:P_comp}
	\Delta_{\rm 21}^2 = T_0^2\, \braket{x_{\rm HI}}^2\, \braket{\eta}^2\, 
					(\Delta^2_{x} + \Delta^2_{\eta} + \Delta^2_{\rho+v} + \cdots ). 
\end{equation}
Again, we omit the higher-order fluctuation terms. These higher-order terms are not always negligible \citep{Lidz07,Georgiev22}, but their omission does not significantly impact the general trends investigated here.
Fig. \ref{fig:ps_zev_comp} shows the cross- and auto-power spectra for the late heating (left), fiducial (middle), and early heating (right) models. The first three terms in Eq.\ref{eq:Px_comp} and Eq.\ref{eq:P_comp} are indicated by the blue dashed, orange dashed-dotted, and green dotted lines, respectively.

At very high redshifts, when there is little heating but Wouthuysen-Field (WF) coupling has already commenced, the spin temperature first decreases (approaching the gas temperature which is initially cooler than the CMB temperature) in overdense regions and the 21 cm signal becomes more negative there. 
This would give a negative correlation between the 21 cm and galaxies. In the redshift range under consideration, we find that the WF coupling is strong enough -- although it does not achieve complete saturation across the entire simulation box -- and so this effect is not observed. 
The marginal impact of this effect is further confirmed by the fact that the $\eta$-galaxy term  ($\Delta^2_{\eta, \rm gal}$; orange dashed-dotted lines) is already positive at high redshift in the top panels of Fig.\ref{fig:ps_zev_comp}.

After that, when the spin temperature is well-coupled to the gas temperature throughout the universe, yet there is still negligible heating, there remains a negative correlation between 21 cm and galaxies. This is because the overdense regions have more neutral hydrogen and hence give more negative 21 cm signals. This is the origin of the negative cross-power signals observed at $z \gtrsim 11$ in the late heating model. One can see that this negative signal is driven by the density contribution ($\Delta^2_{\rho + v, \rm gal}$; green dotted lines) in Fig.\ref{fig:ps_zev_comp}. A similar effect is expected to occur in the fiducial and early heating models at higher redshift than those considered in this study.

Subsequently, X-ray heating starts.
The overdense regions are heated up first and become brighter than typical regions. This leads to the positive cross-correlations at $6.5 \lesssim z \lesssim 11$ in the late heating model, at $z \gtrsim 8$ in the fiducial model, and at $z \gtrsim 9$ in the early heating model, dominated by the temperature term ($\Delta^2_{\eta,\rm gal}$; orange dashed-dotted lines).
After that, we observe negative cross-power spectra in all scenarios. In this study, we particularly focus on this positive-to-negative transition in the cross-power spectrum.\footnote{We note that this transition is  different from the so-called {\it turnover} in the cross-power spectrum. The turnover scale is defined at each redshift as the scale at which the cross-correlation goes from negative to positive (or zero, depending on the small-scale conditions) and is considered to trace the typical size of ionized bubbles around galaxies \citep{Furlanetto07,Lidz09}. This typically happens at much smaller scales than those considered in this paper.} 
We denote the redshift where this transition happens as $z_{\rm tran}$ in the following.

In the fiducial and early heating models, after regions are sufficiently heated, the overdense regions start to become ionized and so dimmer in 21 cm than typical regions, leading to the negative correlation at lower redshifts, dominated by the neutral fraction term ($\Delta^2_{x, \rm gal}$; blue dashed lines).
The positive-to-negative transition happens when the dominant components switch from $\eta$ to $x_{\rm HI}$. We observe $z_{\rm tran} = 8.1$  and 8.9 in the fiducial and early heating models, respectively.

In the late heating model, the X-ray heating and ionization processes overlap significantly in redshift, and thus the situation is slightly different from in the other two models.
As one can see in Fig.\ref{fig:ps_zev_comp}, the neutral fraction field (blue dashed line) contributes positively to the cross-power spectrum when the spin temperature is smaller than the CMB temperature.
Thus even when its contribution becomes comparable to that of spin temperature, the cross-power spectrum is observed as positive.
The neutral fraction fluctuations contribute negatively only after $\braket{\eta}$ changes from negative to positive, i.e., after the so-called heating transition \citep[e.g.,][]{Pritchard12,Fialkov14}. The transition in the late heating model is observed at $z_{\rm tran} = 6.6$.

The distinctions between the models are also illustrated by the inset maps in Fig.\ref{fig:ps_zev_comp},
which showcase the 21 cm signal distributions (red: emission, blue: absorption) along with the locations of the massive haloes (black dots) just before and after the transition redshifts.
While those of the fiducial and early heating models indicate that the transition happens after the large-scale positive clustering disappears due to the X-ray heating and/or before the ionized regions expand,
in the late heating model, the transition happens just when the 21 cm signals outside the ionized regions switch from absorption to emission almost simultaneously.

In reality, even after the end of reionization, neutral hydrogen would remain in galaxies and dense clumps such as damped Ly$\alpha$ systems. Our current method does not take into account this effect, but the 21 cm signal during this post-reionization epoch is expected to be positively correlated with galaxies, leading to yet another sign-change in the cross-power spectrum. 
Indeed, such positive cross-power signals have already been detected at lower redshift \citep{Chang10,Masui13,Anderson18,Wolz22}.

In summary, we generally expect the cross-power spectrum to change sign three times. 
The second, positive-to-negative transition in the cross-power spectrum has been pointed out in previous studies under the assumption of negligible spin temperature fluctuations \citep[e.g.][]{Lidz09}\footnote{Related discussion in the context of the 21 cm auto-power spectrum are also found in some early studies \citep{Furlanetto04,Lidz07}}. The neglect of spin temperature fluctuations in these earlier works led them to find a higher transition redshift than in our current study.

Now let us briefly examine the association with the auto-power spectrum.
The places where the cross-power spectrum goes to zero on large scales are epochs when the large-scale 21 cm distribution becomes nearly uniform, i.e., in such cases the overdense regions are at similar brightness temperatures to typical regions in the universe. 
During these periods, the amplitude of the 21 cm auto-power spectra hence also drops.
The drops (troughs) of the auto-power spectra are seen at $z \sim 11$ and $\sim 6.5$ (late), $z \sim 8$ (fiducial), $z \sim 9$ (early).
This aligns with prior research, which has shown that the large-scale 21 cm auto-power spectrum generally manifests three peaks corresponding to WF, temperature, and ionization fluctuation from high to low redshifts, or exhibits two peaks without a clear distinction between the first two \citep[e.g.,][]{Pritchard07,Pritchard08,Fialkov14,Cohen18,Park19}. 

Observationally, compared to the troughs in the 21 cm auto-power spectrum, the sign changes in the cross-power spectrum could be easier to identify. This is because the cross-power does not require continuous measurements with fine redshift sampling; instead, one just needs detections of positive and negative signals at two different redshifts.
Additionally, the systematic concerns from foreground contamination are reduced in the cross-correlation analysis.
An alternative approach discussed in the literature is to use the 21 cm bispectrum, which is also sensitive to some of the sign change effects studied here \citep{Kamran21}.
In general, the positive-to-negative transition of the cross-power spectrum could serve as an important probe for distinguishing different phases of the EoR and in understanding the nature of the EoR galaxies.

\subsection{Scale dependence}

\begin{figure}
	\includegraphics[width=8cm]{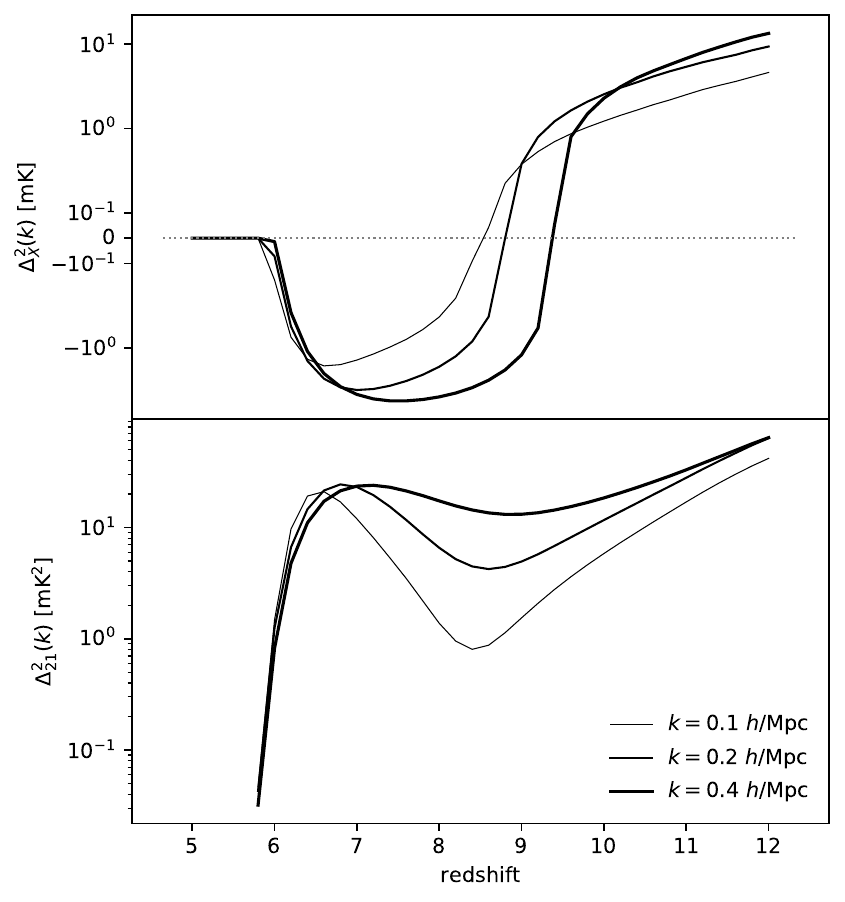}
    \caption{The redshift evolution of cross-power spectra between 21cm and all of the halos in the simulation (top) 
    and 21cm auto-power spectra (bottom). Power spectra at different scales ($k = 0.1$, 0.2, and $0.4~h~\rm Mpc^{-1}$) for the fiducial model are shown. 
    As discussed in the text, we have confirmed that the transition redshift at different scales also corresponds to troughs in the 21 cm auto-power spectrum.
    }
    \label{fig:ps_zev_scale}
\end{figure}

Here, we briefly investigate the scale dependence of the power spectrum. Fig.\ref{fig:ps_zev_scale} shows the cross- and auto-power spectra at $k = 0.1$, 0.2, and 0.4 ~$h~\rm Mpc^{-1}$ for the fiducial model. 
As the heating and ionization both happen in an inside-out manner, the smaller-scale (larger-$k$) signals evolve earlier,
resulting in an earlier transition from positive to negative.
We also confirm that the positive-to-negative transitions in the cross-power spectra roughly correspond to the troughs in the auto-power spectra irrespective of the scales.

\begin{figure}
\includegraphics[width=8cm]{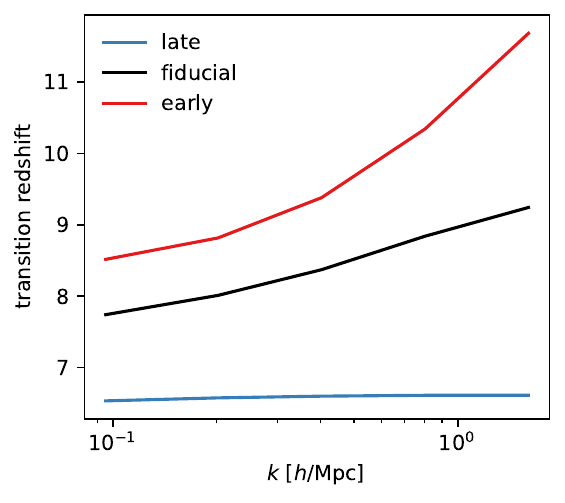}
\caption{The transition redshifts as a function of scale.
In the late heating model, the transition occurs as a result of the global spin temperature average becoming larger than the CMB temperature, and so has no scale dependence. However, in the early and fiducial cases the transition occurs when ionization fluctuations dominate over spin temperature flucations. In these cases, the transition has a scale-dependence.
}
\label{fig:k-ztran}
\end{figure}

Fig.\ref{fig:k-ztran} shows the scale dependence of the transition redshifts for our three models.
Interestingly, while the transition redshifts generally increase as the scale becomes smaller, the late heating model exhibits a flat relation. 
This is because, in the late heating model, the transition is caused solely by the heating transition, not by the inside-out evolution of heating or ionization (see Fig.\ref{fig:ps_zev_comp}).
Such scale (in)dependence, if detected, would provide more information to discern the reionization and heating models than the observation at a single scale.

\subsection{Dependence on galaxy properties}
\label{sec:galpop}

\begin{figure}
\includegraphics[width=8cm]{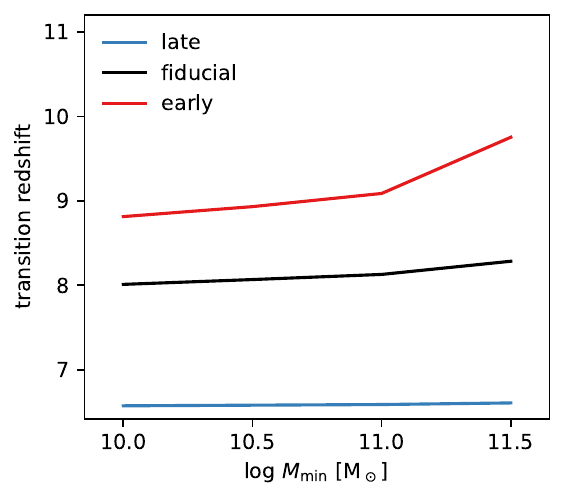}
\caption{The transition redshift measured at $k = 0.2~h~\rm Mpc^{-1}$ as a function of mass limit. 
Galaxy formation, IGM heating, and reionization are accelerated around overdense peaks, and so the transition occurs earlier in the environments of
galaxies residing in more massive and biased halos. 
}
\label{fig:ztran_logMmin}
\end{figure}

Next, we examine the dependence of the transition redshift on the properties of the galaxy populations. 
We first consider the effect of the halo mass limit.
Fig.\ref{fig:ztran_logMmin} shows the transition redshift as a function of the halo mass limit for the three models.
Within the scope of this study, the heating and reionization are primarily driven by massive galaxies; these processes proceed earlier around galaxies residing in larger host halos. This generally results in a higher transition redshift around more massive halos.
While this trend is common across our three models, the dependence on the mass limit is weak in the late heating and fiducial models. 
In fact, the amplitude of the cross-power spectrum itself is boosted by the higher galaxy bias as the minimum mass increases, and thus is more significantly affected by changes in the mass limit. However, the transition redshift is observed to be less dependent on the halo mass limit.
In the early heating model, the stronger dependence may arise from the fact that, at these high redshifts, the ionized bubbles remain isolated. In this case, the bubble size around a galaxy is heavily influenced by its mass and emissivity.

\begin{figure}
\includegraphics[width=8cm]{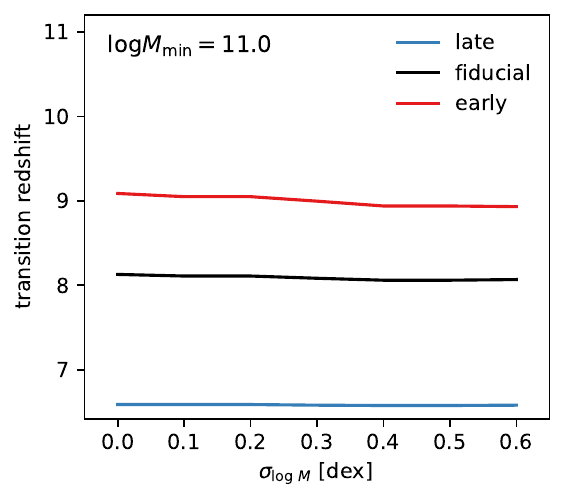}
\caption{The transition redshift measured at $k = 0.2~h~\rm Mpc^{-1}$ as a function of the scatter in the halo mass, $\sigma_{\log M}$. 
We adopt the mass limit of $\log M_{\rm min} = 11.0$.}
\label{fig:ztran_sigma}
\end{figure}

As mentioned in the previous section, there should be some scatter in the halo mass-luminosity relation coming from 
variations in the ISM properties across galaxies residing in halos of a given mass.
Here we crudely account for this by adding Gaussian scatter to each halo mass in our catalog: this changes which halos land above the minimum
threshold mass in our catalog and the total abundance of such halos. This should loosely mimic the effect of scatter in the relationship between the luminosity of an emission line and its host halo mass. We then compute the cross-power spectrum using the resulting galaxy catalog.
The result with added halo mass scatter is presented in Fig.\ref{fig:ztran_sigma}.
Interestingly, the transition redshifts are insensitive to the scatter in the mass-luminosity relation.
The stability of the transition redshifts against both the mass limit and the scatter indicates that we can use different galaxy populations (e.g., those traced by different lines) at different redshifts to measure the transition redshift.

\section{Discussion}
\label{sec:discussion}

Here we discuss the prospects for detecting the cross-power spectra in each of our three 21 cm models. 
Our discussion initially focuses on the halo mass limit, and later we consider the corresponding flux limits for different emission lines. 

\subsection{Detectability}
\label{sec:detectability}

\begin{figure}
\includegraphics[width=8cm]{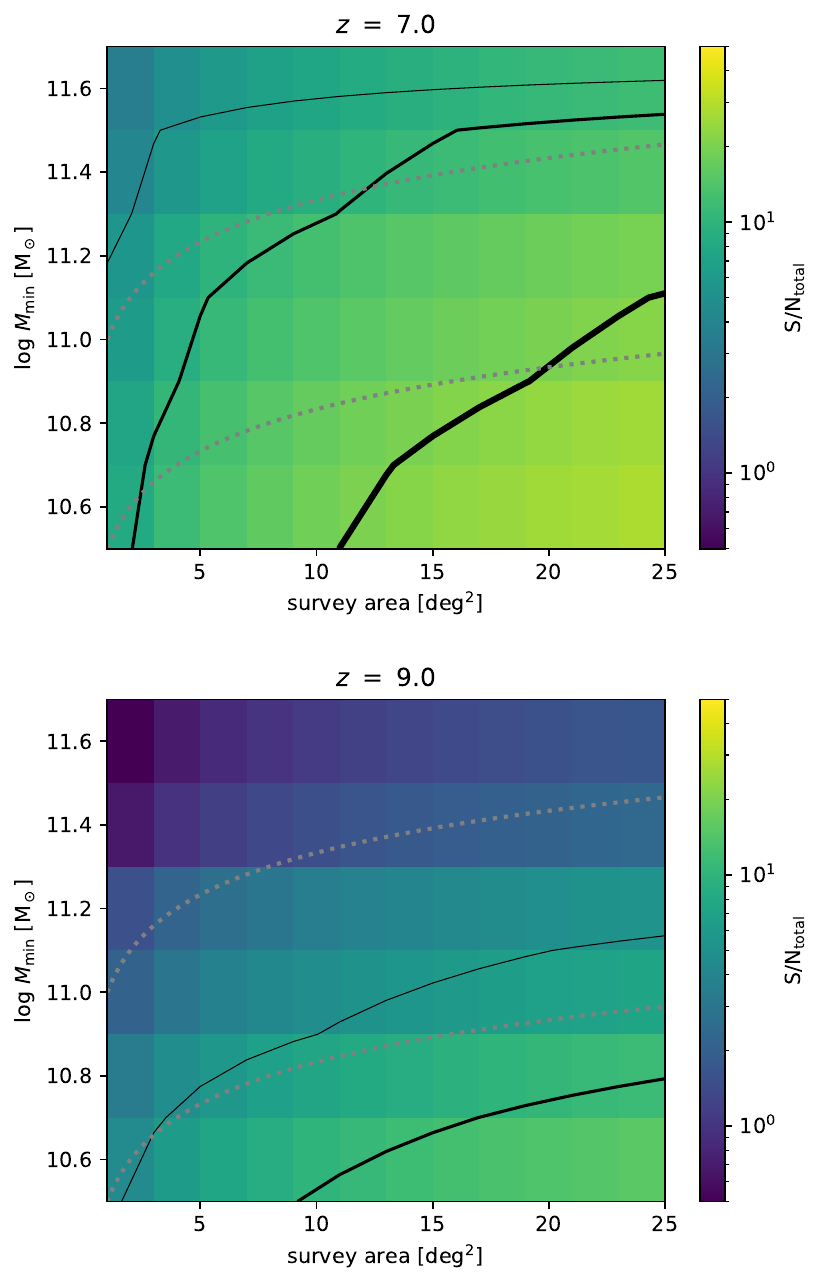}
\caption{The total signal-to-noise ratio (Eq.\ref{eq:sn}) for different survey areas and detection limits ($\log M_{\rm min}$) in the fiducial model. The solid contours correspond to S/N = 5, 10, and 20. The gray dotted lines indicate equal survey times.}
\label{fig:SN_z7_z9}
\end{figure}

The variance of the cross-power spectrum is given by 
\begin{equation} \label{eq:variance}
    \sigma^2_{21, \rm gal}(k,\mu) = \frac{1}{2} [P^2_{21,n}(k,\mu) + \sigma_{21}(k,\mu)\sigma_{\rm gal}(k,\mu)],
\end{equation}
where $\mu$ is the cosine of the angle between wave vector $\bm k$ and the line of sight
and 
\begin{align}
    \sigma_{21}(k,\mu)  &=  P_{21}(k,\mu) + P_{\rm N, 21}(k,\mu), \label{eq:err_21}\\
    \sigma_{\rm gal} (k,\mu) &= P_n (k,\mu) + P_{{\rm N}, \rm gal}(k,\mu), \label{eq:err_gal}
\end{align}
are the variances of the auto-power spectra. 
The noise terms $P_{\rm N, 21}$ and $P_{\rm N, gal}$ take into account the thermal noise in a 21 cm interferometric observation, and shot-noise and redshift uncertainty in a galaxy survey.
For more details, see, for example, Appendix B of \citet{Weinberger20}.
We adopt the same parameters for 21 cm observation (i.e., system temperature and array configuration for upcoming SKA observation) as in \citet{Kubota18} and set the galaxy redshift uncertainty to $\sigma_z = 0.01$.

Although the foregrounds have little correlation with the EoR galaxies, as their amplitude are quite large compared to the EoR 21 cm signals,
one needs to remove or avoid the foreground to detect the cross-power spectrum. That is, residual foregrounds will still contribute to the variance of a cross-power spectrum estimate (even if they produce little average bias).
The foregrounds, which are expected to be spectrally smooth, are most significant in small-$k_{||}$ regions in the two-dimensional Fourier space.
Here we consider wedge avoidance. 
The wedge is given as \citep[see, e.g., Eq. 166 of][]{Liu20} 
\begin{equation}
    k_{||} < k_{||, \rm wedge} \equiv \text{max}\Big[ \frac{H_0 D_c E(z) \theta_0}{c(1+z)}\, k_{\perp}, \, k_{||,\rm min}\Big],
\end{equation}
where we adopt $k_{||,\rm min} = 0.07~h~\rm Mpc^{-1}$ for the flat part of the wedge. 
We assume that the foregrounds above the so-called primary beam wedge -- defined with $\theta_0$ being the size of the primary beam --
can be subtracted and do not affect the resulting signal-to-noise ratio.
With an expected effective area of the SKA antenna
\begin{equation}
    A_e = 462 \Big(\frac{1+z}{9}\Big)^2~\rm [m^2],
\end{equation} 
the size of the primary beam is 
\begin{equation}
    \theta_0 = \frac{\lambda_{\rm 21, obs}}{\sqrt{A_e/\pi}} = 0.16~\rm [radian].
\end{equation}
The angle-averaged noise level is computed via inverse variance weighting.
When avoiding the wedge, one can just assign the modes in the foreground-corrupted wedge infinite variance (i.e., $1/\sigma^2(k,\mu) = 0$).
We thus have
\begin{equation} \label{eq:general_sigma}
    \frac{1}{\sigma^2(k)} =  N_k  \sum_{\mu > k_{||,\rm wedge}/k} \frac{\Delta \mu}{\sigma^2(k,\mu)}, 
\end{equation}
where 
\begin{equation}
    N_k = \frac{\epsilon k^3 V_{\rm surv}}{4\pi^2}
\end{equation}
is the number of modes,
$\epsilon = d\log k$ is the logarithmic bin width, and $V_{\rm surv}$ is the survey volume.

The total signal-to-noise ratio (S/N) is given as 
\begin{equation} \label{eq:sn}
    {\rm S/N}_{\rm total} = \sum_{\rm i} \Big(\frac{P(k_{\rm i})}{\sigma(k_{\rm i})}\Big)^2,
\end{equation}
where the index $\rm i$ runs over all the scale bins $\{k_{\rm i}\}$. 
We adopt a bandwidth of 8 MHz and compute the total S/N with varying the survey area and mass limit.
Fig.\ref{fig:SN_z7_z9} shows the result for the fiducial model at $z = 7$ (top) and $z = 9$ (bottom).
We set the maximum survey area to the planned survey area of SKA, 25 $\rm deg^2$.
The dashed, solid, and broad solid lines show a S/N = 5, 10, and 20, respectively.

From Eq.\ref{eq:sfr_21cmfast}, for an emission line with luminosity proportional to SFR,
the flux limit scales with halo mass as $F_{\rm min} \propto M_{\rm h, min}^{1.5}$ for our choice of $\alpha_* = 0.5$.
The flux limit is inversely proportional to the square root of the observing time $t_{\rm total}$ per unit area $A$: $F_{\rm min} \propto (t_{\rm total}/A)^{-1/2}$. 
For a fixed observing time $t_{\rm total} = \rm const.$, we then have $A \propto F_{\rm min}^2 \propto M_{\rm h, min}^3$.
The gray dotted lines in Fig.\ref{fig:SN_z7_z9} represent arbitrary constant observing times calculated in this way.
One can see that it is better to survey a larger area even at the expense of the flux limit at $z = 7$, which aligns with previous studies \citep{Kubota18, Weinberger20}.
On the other hand, at $z = 9$, the S/N depends only weakly on the choice of these parameters. This is because the number of bright objects decreases strongly towards high redshift, and the shot noise becomes important. 
For reference, we find that at $z = 9$, the shot noise (the second term in Eq.\ref{eq:err_gal}) is $\sim 10$ times larger than the sample variance contribution (the first term in Eq.\ref{eq:err_gal}) when $\log M_{\rm min} = 11.0$, while they are comparable at $z = 7$.

\begin{figure*}
    \centering
    \begin{minipage}[b]{0.45\textwidth}
        \includegraphics[width=\textwidth]{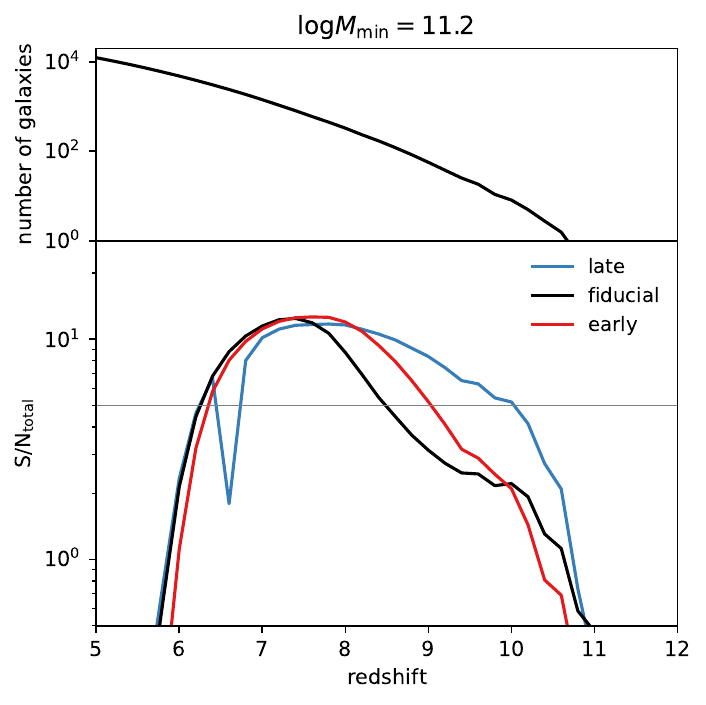}
    \end{minipage}
    \begin{minipage}[b]{0.45\textwidth}
        \includegraphics[width=\textwidth]{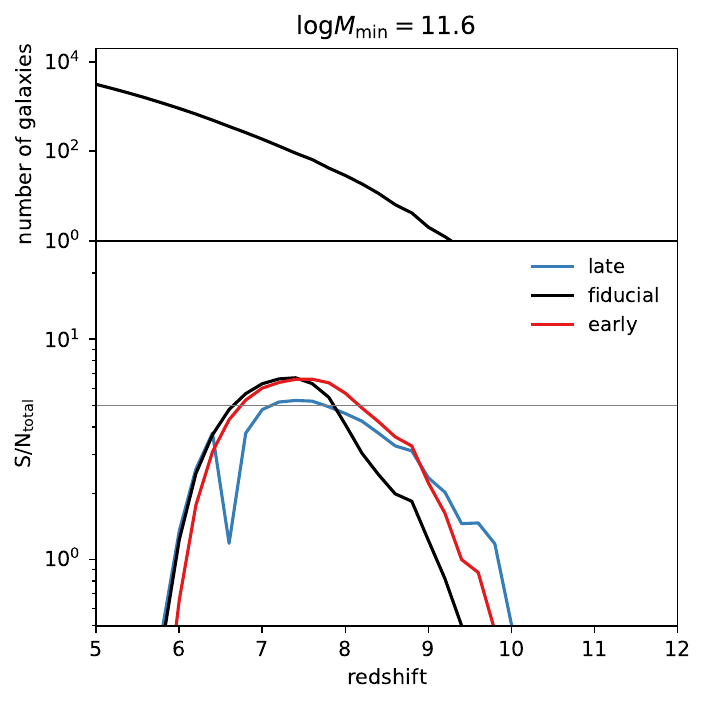}        
    \end{minipage}
\caption{The redshift evolution of the total signal-to-noise ratio (Eq.\ref{eq:sn}) for  a survey volume corresponding to 8 MHz $\times~ 5 \rm ~deg^2$ and a mass limit of $\log M_{\rm min} = 11.2$ (left) and 11.6 (right). The horizontal lines indicate S/N = 5. 
The number of galaxies within the survey volume are shown in the top panels for each value of $\log M_{\rm min}$.
}
\label{fig:zev_SN}
\end{figure*}

Once the observational set-up is fixed, the redshift dependence of the S/N can be investigated.
Fig.\ref{fig:zev_SN} shows the total S/N as a function of redshift for the three models. We adopt a survey area of 5 $\rm deg^2$ and mass limit of $\log M_{\rm min} = 11.2$ (left) and 11.6 (right).
The horizontal lines indicate S/N = 5.
The cross-power spectra at a wide range of redshifts are above this criterion for $\log M_{\rm min} = 11.2$. A detection in the interesting redshift regime of $z=7-8$, where the positive-to-negative transition might occur, is still feasible even with a higher mass limit of $\log M_{\rm min} = 11.6$. 
We note that the sharp drop in the $\rm S/N_{\rm total}$ for the late heating model close to the transition redshift ($z \sim 6.5$) occurs because all modes approach zero simultaneously (see Fig.\ref{fig:k-ztran}).
In most scenarios, the peak S/N is reached near $z = 6-8$, followed by a continuous decrease at higher redshifts. This is different from the auto-power spectrum, where the evolution of the S/N typically shows three peaks, corresponding to signal enhancements from the WF, temperature, and ionization field fluctuations.
In the case of the cross-power spectrum, the decrease in the number of detectable galaxies (i.e., the increase of shot noise) leads to a reduction in the S/N at $z > 8$.

\begin{figure*}
\includegraphics[width=15cm]{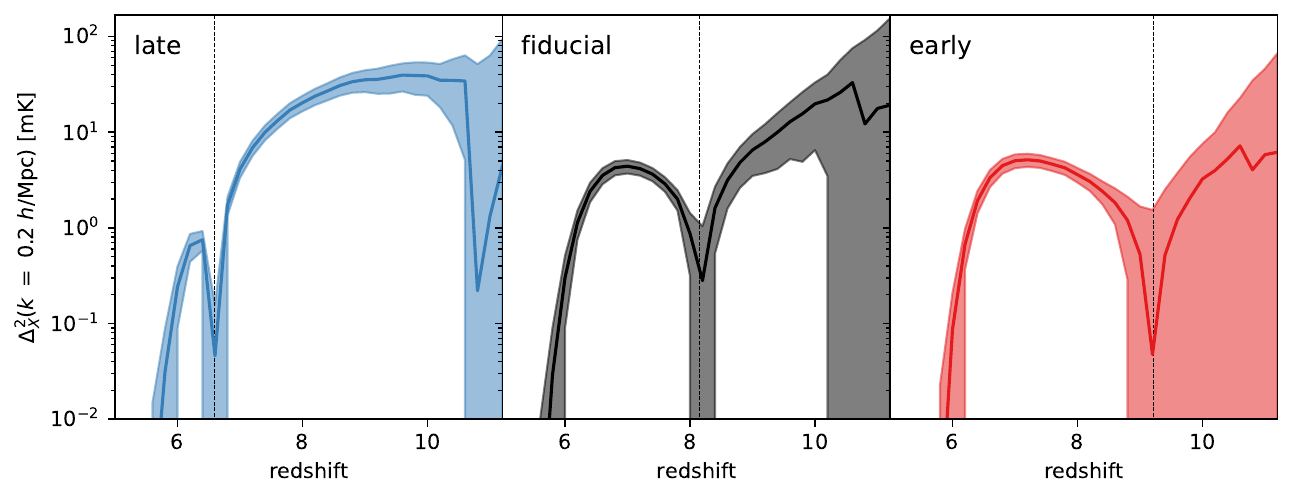}
\caption{The cross-power spectrum amplitude versus redshift at $k = 0.2~h~\rm Mpc^{-1}$ with error bar forecasts. The errors are computed for a survey area of $5~\rm deg^2$ and mass limit $\rm \log M_{\rm min} = 11.2$. The vertical lines show the transition redshifts.}
\label{fig:Px_error}
\end{figure*}

As discussed previously, an important scientific goal is to detect the positive-to-negative transition in the large-scale cross-power spectrum signals. 
Fig.\ref{fig:Px_error} shows the absolute amplitudes of the cross-power spectra at $k = 0.2h~\rm Mpc^{-1}$ for a survey area of 5 $\rm deg^2$ and mass limit of $\log M_{\rm min} = 11.2$. 
The error bar forecasts are shown by the shaded regions, and the transition redshifts are indicated by the vertical lines.
Except for the models where the heating occurs early (rightmost), the signals around the transition redshifts can be detected, suggesting that the transition redshifts can be measured with an uncertainty of $\Delta z \sim 0.2$.

\subsection{Flux limit}
\label{sec:flux_limit}

So far, we have considered galaxy populations in terms of their halo mass. 
Here we discuss the corresponding line fluxes. 
We consider four emission lines: two optical lines, $\rm H\alpha$ and [OIII]5007\AA, and two far-infrared (FIR) lines, [OIII]88$\rm \mu$m and [CII]158$\rm \mu$m.
In the following discussion, we assume $f_{\rm esc}=0$ and ignore dust attenuation for simplicity.
Under our assumption of a sufficiently small values, the escape fraction has a limited impact on the results. 
Below, we describe our method to model the coefficients in the SFR-luminosity relation in Eq.\ref{eq:L-SFR}.

For $\rm H\alpha$, we adopt a value from \citet{Kennicutt98}, 
\begin{equation}
    C_{\rm  H\alpha} = 1.3 \times 10^{41}~\rm (erg/s)/(\rm M_\odot/yr).
\end{equation}
For the oxygen lines, [OIII]5007\AA~and [OIII]88$\rm \mu$m, we follow \citet{Moriwaki18}, where the coefficients are computed as a function of gas metallicity $Z$, density $n$, and ionization parameter $U$ using the photo-ionization calculation code {\sc cloudy} \citep{Ferland17}: 
\begin{equation}
    C_{\rm [OIII]} = C_{\rm [OIII]}(Z, n, U).
\end{equation} 
The ISM properties of high-redshift galaxies can differ from local galaxies. 
We model the galaxy's metallicity using the hydrodynamics simulation IllustrisTNG \citep[TNG300-1;][]{Nelson19}. Within the redshift range focused on here, we find that there exists an almost linear correlation between SFR and metallicity at $\rm SFR \gtrsim 1~\rm M_\odot /yr$ (i.e., $M_{\rm h}\gtrsim 10^{11}~\rm M_\odot$), 
\begin{equation}
    \log Z = \alpha \log {\rm SFR} + \beta, 
\end{equation}
where we find the best-fit parameters are $\alpha = 0.279$ (0.193) and $\beta = -2.784$ (-2.907) for $z = 7$ (9).
The other ISM parameters, namely density and ionization parameter, are difficult to model even with the highest-resolution cosmological hydrodynamic simulations. 
For the density, we adopt a typical value in local star-forming regions, $100~\rm cm^{-3}$. 
As for the ionization parameter, there are several observational indications that high-redshift galaxies are likely to have higher values ($U \sim 10^{-2}$) than local galaxies ($U \sim 10^{-3}$) \citep{Nakajima13,Moriwaki18,Harikane20}. 
While recent studies using zoom-in simulations have predicted a wide range of possible values spanning from $U\sim 10^{-5}$ to 1 \citep{Kohandel23,Nakazato23}, 
it is still reasonable to consider a typical value within the range of $U = 10^{-3} - 10^{-2}$.
We thus show the results for two different ionization parameters: $U = 10^{-2}$ (high $U$) and $U = 10^{-3}$ (low $U$).

In the case of the [CII]158$\rm \mu$m line, we adopt an empirical result obtained in local observations \citep{DeLooze14}:\footnote{We adopt the coefficient for the ``entire literature sample'' in \citet{DeLooze14}. They fit both the coefficient and power in the L-SFR relation and find that the power of SFR is $1.01 \pm 0.02$, consistent with our assumption that the power is 1 in Eq.\ref{eq:L-SFR}.}
\begin{equation} \label{eq:C_CII}
C_{\rm [CII]} = 3.7 \times 10^{40}~\rm (erg/s)/(\rm M_\odot/yr),
\end{equation}
We note, however, that this relation is controversial. While it is found that the SFR-[CII] relation at $4 < z < 6$ is in good agreement with Eq.\ref{eq:C_CII} \citep{Schaerer20},
some studies have pointed out that there is a systematic deviation at $z > 8$, where some galaxies reside beneath the local relation \citep[e.g.,][]{Knudsen16,Laporte19}.
This behavior might be explained by either
a high ionization parameter, low covering fraction of the photodissociation regions, or a combination of both \citep{Harikane20}.
On the other hand, other studies argue that the deviation from the local relation is not that significant \citep[e.g.,][]{Carniani20}.
Therefore, current results regarding [CII]158$\rm \mu$m emission should be approached with caution.

\begin{figure}
\includegraphics[width=8cm]{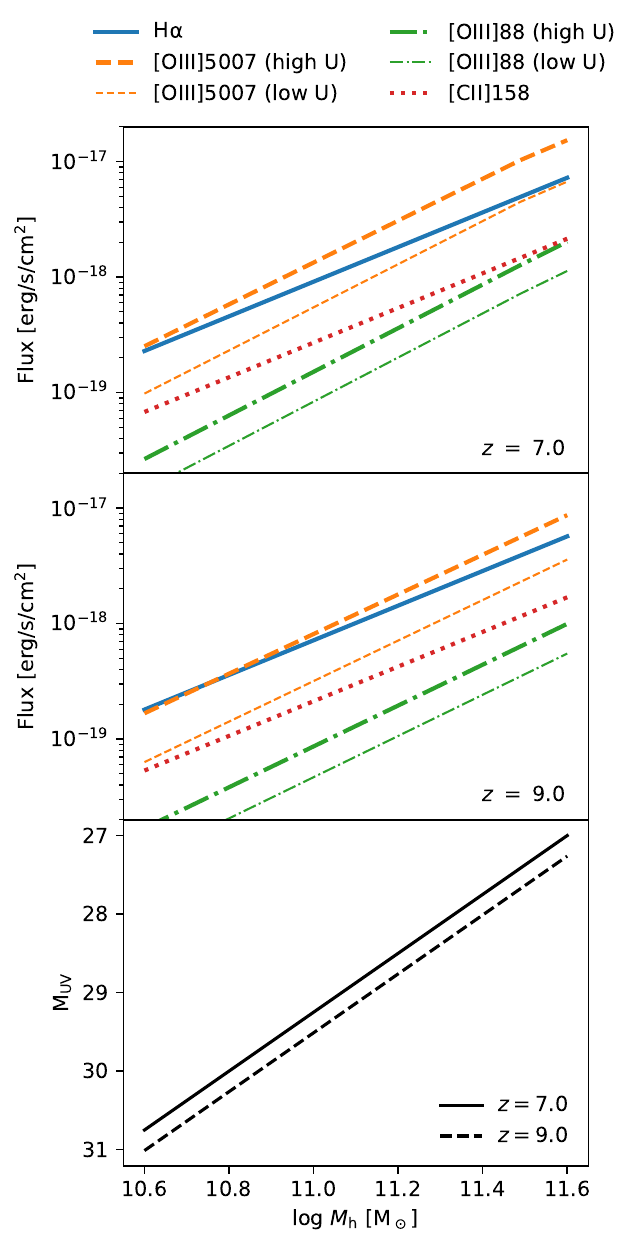}
\caption{The flux limit corresponding to the halo mass threshold used in our analysis at $z = 7$ (top) and $z = 9$ (middle). We compute four different lines $\rm H\alpha$, [OIII] 5007 \AA, [OIII] 88 $\rm \mu m$, and [CII] 158 $\rm \mu m$. For the oxygen lines, we adopt two models with high ($U = 10^{-2}$) and low ($U = 10^{-3}$) ionization parameters.
The bottom panel shows the apparent UV magnitudes at these redshifts.}
\label{fig:mass-flux}
\end{figure}

Fig.\ref{fig:mass-flux} presents the resulting lookup table that shows the mass and flux relations at $z = 7$ (top) and $z = 9$ (middle).
For the oxygen lines, the slope is slightly steeper than the other lines because of the dependence on the metallicity, which scales with the SFR and thus halo mass.
We find that a higher ionization parameter leads to a $\sim$ 0.3 dex increase in the required flux limit. This means that this is a crucial parameter in assessing the detectability of the high-redshift [OIII] emitters and therefore the 21 cm-[OIII] emitter cross-power spectrum. 

From the discussion so far, the maximum mass limit required to detect the signals at $z \sim 6-8$ is $\log M_{\rm h,min} \sim 11.6$ (see the right panel of Fig.\ref{fig:zev_SN}). 
This corresponds to flux limits of $\sim 10^{-17}~\rm erg/s/cm^2$ for optical lines and $\sim 10^{-18}~\rm erg/s/cm^2$ for FIR lines.
To conduct a blind galaxy survey with the required sensitivities, the use of a new telescope may be necessary. 
For instance, the Large Submillimeter Telescope \citep[LST, 2030s-;][]{Kawabe16} will have a large 0.5 $\rm deg^2$ field-of-view and a spectral resolution of $R \sim 1000$ and is one possibility.
In its proposed long-term program spanning a few years with a total observing time of 9,000 hours, the telescope is expected to achieve a 5-$\sigma$ sensitivity of $\sim$ 0.1 Jy~km/s for [OIII]88$\rm \mu$m at $z \gtrsim 8$, equivalent to $\sim 10^{-18}~\rm erg/s/cm^2$, over 2 $\rm deg^2$. 
While our predictions indicate that the S/N is likely to remain at $\sim 2-3$ even with such observations, a signal at this level would still have the potential to yield important scientific information.

It may be difficult to conduct a blind spectroscopic survey with much higher sensitivity to detect the transition redshift.
For this purpose, follow-up observations of galaxies already detected by deep photometric surveys could be a crucial strategy.
To predict the required depth for such surveys, we compute the UV luminosity from the SFR as 
\begin{align}
    \frac{\rm SFR}{\rm M_\odot/yr} = \mathcal{K}_{UV} \,\frac{L_\nu(\rm UV)}{\rm erg/s/Hz},
\end{align}
where we adopt $\mathcal{K}_{\rm UV} = 10^{-28}$ assuming that the galaxies have sub-solar metallicities at these high redshifts \citep{Madau14}.
The bottom panel of Fig.\ref{fig:mass-flux} shows the apparent UV magnitudes. 
For instance, the Nancy Grace Roman Space Telescope \citep{Spergel15} is capable of detecting $z > 5$ Lyman-break galaxies (LBGs). 
Its High Latitude Surveys \citep{Dore18,Wang22} are proposed to achieve NIR (rest-frame UV at the EoR) magnitude $m_{\rm AB} <26.5$ over $\sim 100~\rm deg^2$, and \citet{LaPlante23} have already discussed the possibility of using them for cross-correlation analyses. 
If a much deeper observation with $m_{\rm AB} < 28.5$ is available for a smaller area overlapping with the SKA sky coverage, it could reach $\log M_{\rm min} = 11.2$.
After selecting photometric samples with such an imaging survey (for e.g., a few hundred galaxies; see the top panels of Fig.\ref{fig:zev_SN}), it is expected to be relatively easy to detect their emission lines if we use JWST.

\subsection{Future prospects}

As we have seen, cross-power spectrum measurements may be crucial not only for mitigating foreground contamination systematics, but also for extracting information about the reionization process via measurements of 
the transition redshift and the associated redshift evolution. 
These can be used for disentangling degeneracies in 21cm auto-power spectra observations, e.g., the degeneracy between the ``density-driven'' and ``reionization-deiven'' models discussed in \citet{HERA22}.
In particular, as models for the sources of reionization are constrained to some extent by other observations such as UV luminosity functions and CMB, 
the cross-correlation observations should provide powerful constraints on heating models. 
For this purpose, it is crucial to observe relatively high redshifts, $z = 8-10$.
In this paper, we considered only three representative 21 cm models, but our methodology can be employed across a wider range of models in future resarch.

In this study, we employed a simplified galaxy model which can be refined in future studies. 
For example, we assume that most of the physical parameters are redshift-independent.
In reality, however, the emissivity of high-redshift galaxies and other sources will vary with redshift and other factors.
For instance, population III stars may have a harder spectrum, leading to enhanced ionizing capability \citep[e.g.,][]{Tanaka21}. 
Also, the stochasticity of the star formation is thought to be more prominent at higher redshift \citep[e.g.,][]{Ciesla23}. This scatter in the relationship between SFR and halo mass might impact the 21 cm fluctuation signal and the detectability of galaxies \citep{Reis22}.

Another important factor we have neglected is the presence of residual foreground contamination leftover after wedge-filtering. Residual foregrounds might degrade our S/N forecasts, while details of the foreground removal or avoidance procedure might also further impact the cross-power spectrum measurements themselves \citep{Yoshiura18}.
If residual foreground noise is important, this might necessitate an even deeper galaxy survey (i.e., deeper than $m_{\rm AB} \sim 29$) to offset
the increased 21 cm signal variance. 
More detailed investigations taking into account plausible levels of residual foreground noise should be done in future studies.

Since the biggest challenge is detecting faint high-redshift galaxies over a wide area, one may need to carefully consider the analytical and observational strategy.
For instance, instead of using the galaxy number density, one could employ the luminosity-weighted number density for cross-correlation with 21 cm fields. It is known that certain types of weighting enhance the S/N of galaxy power spectrum measurements \citep{Seljak09, Hamaus10, Cai11}.
The weighting does not necessarily have to be linear with respect to luminosity, 
it would be interesting to determine the optimal weighting strategy in future work.
Since the weighting would affect not only the S/N but also the amplitude and scale-dependence of the cross-power spectrum, applying different weighting methods, which can be tested at no additional cost, could potentially extract more information.

Another interesting alternative approach involves line intensity mapping, which measures the large-scale fluctuations of integrated line emissions rather than detecting individual galaxies \citep{Kovetz17}. Such an option has been partly studied in previous studies \citep[e.g.,][]{Lidz11,Gong12,Silva13,Heneka21,Moriwaki19}.
There are a few plans that could measure lines from galaxies at the redshifts of interest here \citep[e.g., CDIM][]{Cooray19}.
We expect all the general trends discussed in this study to apply to the cross-power spectrum with line intensity mapping as well, but this should be confirmed in future studies.
Modeling the line intensity mapping signals will require accounting for the role of low-luminosity galaxies.
The scatter in the SFR-halo mass relation could also have a large impact on the line intensity power spectrum \citep{Murmu23}.

\section{Conclusion}

In this study, we examined the redshift evolution of the cross-power spectra between emission line galaxies and the 21~cm field for the first time. 
We used {\sc 21cmFAST} to generate 21 cm brightness temperature fields and considered three different heating models.
We adopted a simple model for the galaxy populations, which allowed us to investigate the general evolutionary trends in the cross-power spectrum and
the dependence on the emission lines used.
Our findings are summarized as follows:
\begin{enumerate}

\item The positive-to-negative transition in the cross-power spectrum corresponds to a trough in the 21 cm auto-power spectrum. 
As the {\it pre-} and {\it post}-transition signals clearly differ in their signs, the transition is easier to identify in the cross-power spectrum.

\item The transition occurs when both 1.) the global spin temperature rises above the CMB temperature, and 2.) the ionization fluctuations dominate over spin temperature fluctuations. If the first condition occurs after the second one, the transition is scale-independent. If the second condition is reached after the first one, the transition has a scale dependence. Measuring the scale dependence of the transition redshift thus directly constrains important transitions in the 21~cm field.

\item The signal-to-noise ratio of the cross-power spectrum peaks at around $z = 7-8$ in the models considered, due to the rapid decline in the abundance of detectable galaxies at high redshifts.
In scenarios with late X-ray heating, the transition redshift can be detected with a combination of an SKA-like 21 cm observation and a galaxy survey that can achieve sensitivities of $\sim 10^{-18}~\rm erg/s/cm^2$ for rest-frame optical lines and $\sim 10^{-19}~\rm erg/s/cm^2$ for rest-frame FIR lines. To achieve small flux limits, imaging surveys conducting a pre-selection for high-redshift candidates may be important.

\end{enumerate}

Using the methodology adopted in this paper, we plan to explore a wider range of reionization and heating models in future studies.

\section*{Acknowledgements}

We thank
Kotaro Kohno for giving insightful comments on galaxy observations,
and Shintaro Yoshiura and Hayato Shimabukuro for discussion on 21 cm observations.
This work was initiated at "the S\~{a}o Paulo School of Advanced Science on First Light: Stars, galaxies and black holes in the epoch of reionization", sponsored by FAPESP.
KM acknowledges JSPS KAKENHI Grant Number 23K03446 and 23K20035.
AL acknowledges support through NASA ATP grant 80NSSC20K0497.

\section*{Data Availability}
We will share the simulated data upon reasonable request.



\bibliographystyle{mnras}
\bibliography{bibtex_library} % if your bibtex file is called example.bib

\begin{thebibliography}{}
\makeatletter
\relax
\def\mn@urlcharsother{\let\do\@makeother \do\$\do\&\do\#\do\^\do\_\do\%\do\~}
\def\mn@doi{\begingroup\mn@urlcharsother \@ifnextchar [ {\mn@doi@}
  {\mn@doi@[]}}
\def\mn@doi@[#1]#2{\def\@tempa{#1}\ifx\@tempa\@empty \href
  {http://dx.doi.org/#2} {doi:#2}\else \href {http://dx.doi.org/#2} {#1}\fi
  \endgroup}
\def\mn@eprint#1#2{\mn@eprint@#1:#2::\@nil}
\def\mn@eprint@arXiv#1{\href {http://arxiv.org/abs/#1} {{\tt arXiv:#1}}}
\def\mn@eprint@dblp#1{\href {http://dblp.uni-trier.de/rec/bibtex/#1.xml}
  {dblp:#1}}
\def\mn@eprint@#1:#2:#3:#4\@nil{\def\@tempa {#1}\def\@tempb {#2}\def\@tempc
  {#3}\ifx \@tempc \@empty \let \@tempc \@tempb \let \@tempb \@tempa \fi \ifx
  \@tempb \@empty \def\@tempb {arXiv}\fi \@ifundefined
  {mn@eprint@\@tempb}{\@tempb:\@tempc}{\expandafter \expandafter \csname
  mn@eprint@\@tempb\endcsname \expandafter{\@tempc}}}

\bibitem[\protect\citeauthoryear{{Abdurashidova} et~al.,}{{Abdurashidova}
  et~al.}{2022a}]{HERA22b}
{Abdurashidova} Z.,  et~al., 2022a, \mn@doi [\apj] {10.3847/1538-4357/ac2ffc},
  \href {https://ui.adsabs.harvard.edu/abs/2022ApJ...924...51A} {924, 51}

\bibitem[\protect\citeauthoryear{{Abdurashidova} et~al.,}{{Abdurashidova}
  et~al.}{2022b}]{HERA22}
{Abdurashidova} Z.,  et~al., 2022b, \mn@doi [\apj] {10.3847/1538-4357/ac1c78},
  \href {https://ui.adsabs.harvard.edu/abs/2022ApJ...925..221A} {925, 221}

\bibitem[\protect\citeauthoryear{{Anderson} et~al.,}{{Anderson}
  et~al.}{2018}]{Anderson18}
{Anderson} C.~J.,  et~al., 2018, \mn@doi [\mnras] {10.1093/mnras/sty346}, \href
  {https://ui.adsabs.harvard.edu/abs/2018MNRAS.476.3382A} {476, 3382}

\bibitem[\protect\citeauthoryear{{Atek}, {Richard}, {Kneib}  \&
  {Schaerer}}{{Atek} et~al.}{2018}]{Atek18}
{Atek} H.,  {Richard} J.,  {Kneib} J.-P.,   {Schaerer} D.,  2018, \mn@doi
  [\mnras] {10.1093/mnras/sty1820}, \href
  {https://ui.adsabs.harvard.edu/abs/2018MNRAS.479.5184A} {479, 5184}

\bibitem[\protect\citeauthoryear{{Barry} et~al.,}{{Barry}
  et~al.}{2019}]{Barry19}
{Barry} N.,  et~al., 2019, \mn@doi [\apj] {10.3847/1538-4357/ab40a8}, \href
  {https://ui.adsabs.harvard.edu/abs/2019ApJ...884....1B} {884, 1}

\bibitem[\protect\citeauthoryear{{Beane} \& {Lidz}}{{Beane} \&
  {Lidz}}{2018}]{Beane18}
{Beane} A.,  {Lidz} A.,  2018, \mn@doi [\apj] {10.3847/1538-4357/aae388}, \href
  {https://ui.adsabs.harvard.edu/abs/2018ApJ...867...26B} {867, 26}

\bibitem[\protect\citeauthoryear{{Beane}, {Villaescusa-Navarro}  \&
  {Lidz}}{{Beane} et~al.}{2019}]{Beane19}
{Beane} A.,  {Villaescusa-Navarro} F.,   {Lidz} A.,  2019, \mn@doi [\apj]
  {10.3847/1538-4357/ab0a08}, \href
  {https://ui.adsabs.harvard.edu/abs/2019ApJ...874..133B} {874, 133}

\bibitem[\protect\citeauthoryear{{Beardsley} et~al.,}{{Beardsley}
  et~al.}{2016}]{Beardsley16}
{Beardsley} A.~P.,  et~al., 2016, \mn@doi [\apj] {10.3847/1538-4357/833/1/102},
  \href {https://ui.adsabs.harvard.edu/abs/2016ApJ...833..102B} {833, 102}

\bibitem[\protect\citeauthoryear{{Bhatawdekar}, {Conselice},
  {Margalef-Bentabol}  \& {Duncan}}{{Bhatawdekar} et~al.}{2019}]{Bhatawdekar19}
{Bhatawdekar} R.,  {Conselice} C.~J.,  {Margalef-Bentabol} B.,   {Duncan} K.,
  2019, \mn@doi [\mnras] {10.1093/mnras/stz866}, \href
  {https://ui.adsabs.harvard.edu/abs/2019MNRAS.486.3805B} {486, 3805}

\bibitem[\protect\citeauthoryear{{Bouwens} et~al.,}{{Bouwens}
  et~al.}{2015}]{Bouwens15}
{Bouwens} R.~J.,  et~al., 2015, \mn@doi [\apj] {10.1088/0004-637X/803/1/34},
  \href {https://ui.adsabs.harvard.edu/abs/2015ApJ...803...34B} {803, 34}

\bibitem[\protect\citeauthoryear{{Cai}, {Bernstein}  \& {Sheth}}{{Cai}
  et~al.}{2011}]{Cai11}
{Cai} Y.-C.,  {Bernstein} G.,   {Sheth} R.~K.,  2011, \mn@doi [\mnras]
  {10.1111/j.1365-2966.2010.17969.x}, \href
  {https://ui.adsabs.harvard.edu/abs/2011MNRAS.412..995C} {412, 995}

\bibitem[\protect\citeauthoryear{{Carniani} et~al.,}{{Carniani}
  et~al.}{2020}]{Carniani20}
{Carniani} S.,  et~al., 2020, \mn@doi [\mnras] {10.1093/mnras/staa3178}, \href
  {https://ui.adsabs.harvard.edu/abs/2020MNRAS.499.5136C} {499, 5136}

\bibitem[\protect\citeauthoryear{{Chang}, {Pen}, {Bandura}  \&
  {Peterson}}{{Chang} et~al.}{2010}]{Chang10}
{Chang} T.-C.,  {Pen} U.-L.,  {Bandura} K.,   {Peterson} J.~B.,  2010, \mn@doi
  [\nat] {10.1038/nature09187}, \href
  {https://ui.adsabs.harvard.edu/abs/2010Natur.466..463C} {466, 463}

\bibitem[\protect\citeauthoryear{{Ciesla} et~al.,}{{Ciesla}
  et~al.}{2023}]{Ciesla23}
{Ciesla} L.,  et~al., 2023, \mn@doi [arXiv e-prints]
  {10.48550/arXiv.2309.15720}, \href
  {https://ui.adsabs.harvard.edu/abs/2023arXiv230915720C} {p. arXiv:2309.15720}

\bibitem[\protect\citeauthoryear{{Cohen}, {Fialkov}, {Barkana}  \&
  {Lotem}}{{Cohen} et~al.}{2017}]{Cohen17}
{Cohen} A.,  {Fialkov} A.,  {Barkana} R.,   {Lotem} M.,  2017, \mn@doi [\mnras]
  {10.1093/mnras/stx2065}, \href
  {https://ui.adsabs.harvard.edu/abs/2017MNRAS.472.1915C} {472, 1915}

\bibitem[\protect\citeauthoryear{{Cohen}, {Fialkov}  \& {Barkana}}{{Cohen}
  et~al.}{2018}]{Cohen18}
{Cohen} A.,  {Fialkov} A.,   {Barkana} R.,  2018, \mn@doi [\mnras]
  {10.1093/mnras/sty1094}, \href
  {https://ui.adsabs.harvard.edu/abs/2018MNRAS.478.2193C} {478, 2193}

\bibitem[\protect\citeauthoryear{{Cooray} et~al.,}{{Cooray}
  et~al.}{2019}]{Cooray19}
{Cooray} A.,  et~al., 2019, in Bulletin of the American Astronomical Society.
  p.~23 (\mn@eprint {arXiv} {1903.03144}), \mn@doi{10.48550/arXiv.1903.03144}

\bibitem[\protect\citeauthoryear{{Datta}, {Mellema}, {Mao}, {Iliev}, {Shapiro}
  \& {Ahn}}{{Datta} et~al.}{2012}]{Datta12}
{Datta} K.~K.,  {Mellema} G.,  {Mao} Y.,  {Iliev} I.~T.,  {Shapiro} P.~R.,
  {Ahn} K.,  2012, \mn@doi [\mnras] {10.1111/j.1365-2966.2012.21293.x}, \href
  {https://ui.adsabs.harvard.edu/abs/2012MNRAS.424.1877D} {424, 1877}

\bibitem[\protect\citeauthoryear{{De Looze} et~al.,}{{De Looze}
  et~al.}{2014}]{DeLooze14}
{De Looze} I.,  et~al., 2014, \mn@doi [\aap] {10.1051/0004-6361/201322489},
  \href {https://ui.adsabs.harvard.edu/abs/2014A&A...568A..62D} {568, A62}

\bibitem[\protect\citeauthoryear{{Dillon} et~al.,}{{Dillon}
  et~al.}{2014}]{Dillon14}
{Dillon} J.~S.,  et~al., 2014, \mn@doi [\prd] {10.1103/PhysRevD.89.023002},
  \href {https://ui.adsabs.harvard.edu/abs/2014PhRvD..89b3002D} {89, 023002}

\bibitem[\protect\citeauthoryear{{Dor{\'e}} et~al.,}{{Dor{\'e}}
  et~al.}{2018}]{Dore18}
{Dor{\'e}} O.,  et~al., 2018, \mn@doi [arXiv e-prints]
  {10.48550/arXiv.1804.03628}, \href
  {https://ui.adsabs.harvard.edu/abs/2018arXiv180403628D} {p. arXiv:1804.03628}

\bibitem[\protect\citeauthoryear{{Dumitru}, {Kulkarni}, {Lagache}  \&
  {Haehnelt}}{{Dumitru} et~al.}{2019}]{Dumitru19}
{Dumitru} S.,  {Kulkarni} G.,  {Lagache} G.,   {Haehnelt} M.~G.,  2019, \mn@doi
  [\mnras] {10.1093/mnras/stz617}, \href
  {https://ui.adsabs.harvard.edu/abs/2019MNRAS.485.3486D} {485, 3486}

\bibitem[\protect\citeauthoryear{{Eastwood} et~al.,}{{Eastwood}
  et~al.}{2019}]{Eastwood19}
{Eastwood} M.~W.,  et~al., 2019, \mn@doi [\aj] {10.3847/1538-3881/ab2629},
  \href {https://ui.adsabs.harvard.edu/abs/2019AJ....158...84E} {158, 84}

\bibitem[\protect\citeauthoryear{{Ewall-Wice} et~al.,}{{Ewall-Wice}
  et~al.}{2016}]{Ewall-Wice16}
{Ewall-Wice} A.,  et~al., 2016, \mn@doi [\mnras] {10.1093/mnras/stw1022}, \href
  {https://ui.adsabs.harvard.edu/abs/2016MNRAS.460.4320E} {460, 4320}

\bibitem[\protect\citeauthoryear{{Ferland} et~al.,}{{Ferland}
  et~al.}{2017}]{Ferland17}
{Ferland} G.~J.,  et~al., 2017, \rmxaa, \href
  {https://ui.adsabs.harvard.edu/abs/2017RMxAA..53..385F} {53, 385}

\bibitem[\protect\citeauthoryear{{Fialkov} \& {Barkana}}{{Fialkov} \&
  {Barkana}}{2014}]{Fialkov14}
{Fialkov} A.,  {Barkana} R.,  2014, \mn@doi [\mnras] {10.1093/mnras/stu1744},
  \href {https://ui.adsabs.harvard.edu/abs/2014MNRAS.445..213F} {445, 213}

\bibitem[\protect\citeauthoryear{{Finkelstein} et~al.,}{{Finkelstein}
  et~al.}{2015}]{Finkelstein15}
{Finkelstein} S.~L.,  et~al., 2015, \mn@doi [\apj]
  {10.1088/0004-637X/810/1/71}, \href
  {https://ui.adsabs.harvard.edu/abs/2015ApJ...810...71F} {810, 71}

\bibitem[\protect\citeauthoryear{{Furlanetto} \& {Lidz}}{{Furlanetto} \&
  {Lidz}}{2007}]{Furlanetto07}
{Furlanetto} S.~R.,  {Lidz} A.,  2007, \mn@doi [\apj] {10.1086/513009}, \href
  {https://ui.adsabs.harvard.edu/abs/2007ApJ...660.1030F} {660, 1030}

\bibitem[\protect\citeauthoryear{{Furlanetto}, {Zaldarriaga}  \&
  {Hernquist}}{{Furlanetto} et~al.}{2004}]{Furlanetto04}
{Furlanetto} S.~R.,  {Zaldarriaga} M.,   {Hernquist} L.,  2004, \mn@doi [\apj]
  {10.1086/423025}, \href
  {https://ui.adsabs.harvard.edu/abs/2004ApJ...613....1F} {613, 1}

\bibitem[\protect\citeauthoryear{{Gehlot} et~al.,}{{Gehlot}
  et~al.}{2019}]{Gehlot19}
{Gehlot} B.~K.,  et~al., 2019, \mn@doi [\mnras] {10.1093/mnras/stz1937}, \href
  {https://ui.adsabs.harvard.edu/abs/2019MNRAS.488.4271G} {488, 4271}

\bibitem[\protect\citeauthoryear{{Gehlot} et~al.,}{{Gehlot}
  et~al.}{2020}]{Gehlot20}
{Gehlot} B.~K.,  et~al., 2020, \mn@doi [\mnras] {10.1093/mnras/staa3093}, \href
  {https://ui.adsabs.harvard.edu/abs/2020MNRAS.499.4158G} {499, 4158}

\bibitem[\protect\citeauthoryear{{Georgiev}, {Mellema}, {Giri}  \&
  {Mondal}}{{Georgiev} et~al.}{2022}]{Georgiev22}
{Georgiev} I.,  {Mellema} G.,  {Giri} S.~K.,   {Mondal} R.,  2022, \mn@doi
  [\mnras] {10.1093/mnras/stac1230}, \href
  {https://ui.adsabs.harvard.edu/abs/2022MNRAS.513.5109G} {513, 5109}

\bibitem[\protect\citeauthoryear{{Gong}, {Cooray}, {Silva}, {Santos}, {Bock},
  {Bradford}  \& {Zemcov}}{{Gong} et~al.}{2012}]{Gong12}
{Gong} Y.,  {Cooray} A.,  {Silva} M.,  {Santos} M.~G.,  {Bock} J.,  {Bradford}
  C.~M.,   {Zemcov} M.,  2012, \mn@doi [\apj] {10.1088/0004-637X/745/1/49},
  \href {https://ui.adsabs.harvard.edu/abs/2012ApJ...745...49G} {745, 49}

\bibitem[\protect\citeauthoryear{{Greig} \& {Mesinger}}{{Greig} \&
  {Mesinger}}{2018}]{Greig18}
{Greig} B.,  {Mesinger} A.,  2018, \mn@doi [\mnras] {10.1093/mnras/sty796},
  \href {https://ui.adsabs.harvard.edu/abs/2018MNRAS.477.3217G} {477, 3217}

\bibitem[\protect\citeauthoryear{{Hamaus}, {Seljak}, {Desjacques}, {Smith}  \&
  {Baldauf}}{{Hamaus} et~al.}{2010}]{Hamaus10}
{Hamaus} N.,  {Seljak} U.,  {Desjacques} V.,  {Smith} R.~E.,   {Baldauf} T.,
  2010, \mn@doi [\prd] {10.1103/PhysRevD.82.043515}, \href
  {https://ui.adsabs.harvard.edu/abs/2010PhRvD..82d3515H} {82, 043515}

\bibitem[\protect\citeauthoryear{{Harikane} et~al.,}{{Harikane}
  et~al.}{2020}]{Harikane20}
{Harikane} Y.,  et~al., 2020, \mn@doi [\apj] {10.3847/1538-4357/ab94bd}, \href
  {https://ui.adsabs.harvard.edu/abs/2020ApJ...896...93H} {896, 93}

\bibitem[\protect\citeauthoryear{{Harikane}, {Nakajima}, {Ouchi}, {Umeda},
  {Isobe}, {Ono}, {Xu}  \& {Zhang}}{{Harikane} et~al.}{2023}]{Harikane23}
{Harikane} Y.,  {Nakajima} K.,  {Ouchi} M.,  {Umeda} H.,  {Isobe} Y.,  {Ono}
  Y.,  {Xu} Y.,   {Zhang} Y.,  2023, \mn@doi [arXiv e-prints]
  {10.48550/arXiv.2304.06658}, \href
  {https://ui.adsabs.harvard.edu/abs/2023arXiv230406658H} {p. arXiv:2304.06658}

\bibitem[\protect\citeauthoryear{{Heneka} \& {Cooray}}{{Heneka} \&
  {Cooray}}{2021}]{Heneka21}
{Heneka} C.,  {Cooray} A.,  2021, \mn@doi [\mnras] {10.1093/mnras/stab1842},
  \href {https://ui.adsabs.harvard.edu/abs/2021MNRAS.506.1573H} {506, 1573}

\bibitem[\protect\citeauthoryear{{Heneka} \& {Mesinger}}{{Heneka} \&
  {Mesinger}}{2020}]{Heneka20}
{Heneka} C.,  {Mesinger} A.,  2020, \mn@doi [\mnras] {10.1093/mnras/staa1517},
  \href {https://ui.adsabs.harvard.edu/abs/2020MNRAS.496..581H} {496, 581}

\bibitem[\protect\citeauthoryear{{Hutter}, {Dayal}, {M{\"u}ller}  \&
  {Trott}}{{Hutter} et~al.}{2017}]{Hutter17}
{Hutter} A.,  {Dayal} P.,  {M{\"u}ller} V.,   {Trott} C.~M.,  2017, \mn@doi
  [\apj] {10.3847/1538-4357/836/2/176}, \href
  {http://adsabs.harvard.edu/abs/2017ApJ...836..176H} {836, 176}

\bibitem[\protect\citeauthoryear{{Hutter}, {Heneka}, {Dayal}, {Gottl{\"o}ber},
  {Mesinger}, {Trebitsch}  \& {Yepes}}{{Hutter} et~al.}{2023}]{Hutter23}
{Hutter} A.,  {Heneka} C.,  {Dayal} P.,  {Gottl{\"o}ber} S.,  {Mesinger} A.,
  {Trebitsch} M.,   {Yepes} G.,  2023, \mn@doi [arXiv e-prints]
  {10.48550/arXiv.2306.03156}, \href
  {https://ui.adsabs.harvard.edu/abs/2023arXiv230603156H} {p. arXiv:2306.03156}

\bibitem[\protect\citeauthoryear{{Kamran}, {Majumdar}, {Ghara}, {Mellema},
  {Bharadwaj}, {Pritchard}, {Mondal}  \& {Iliev}}{{Kamran}
  et~al.}{2021}]{Kamran21}
{Kamran} M.,  {Majumdar} S.,  {Ghara} R.,  {Mellema} G.,  {Bharadwaj} S.,
  {Pritchard} J.~R.,  {Mondal} R.,   {Iliev} I.~T.,  2021, \mn@doi [arXiv
  e-prints] {10.48550/arXiv.2108.08201}, \href
  {https://ui.adsabs.harvard.edu/abs/2021arXiv210808201K} {p. arXiv:2108.08201}

\bibitem[\protect\citeauthoryear{{Kannan}, {Smith}, {Garaldi}, {Shen},
  {Vogelsberger}, {Pakmor}, {Springel}  \& {Hernquist}}{{Kannan}
  et~al.}{2022}]{Kannan22}
{Kannan} R.,  {Smith} A.,  {Garaldi} E.,  {Shen} X.,  {Vogelsberger} M.,
  {Pakmor} R.,  {Springel} V.,   {Hernquist} L.,  2022, \mn@doi [\mnras]
  {10.1093/mnras/stac1557}, \href
  {https://ui.adsabs.harvard.edu/abs/2022MNRAS.514.3857K} {514, 3857}

\bibitem[\protect\citeauthoryear{{Kashikawa} et~al.,}{{Kashikawa}
  et~al.}{2006}]{Kashikawa06}
{Kashikawa} N.,  et~al., 2006, \mn@doi [\apj] {10.1086/504966}, \href
  {https://ui.adsabs.harvard.edu/abs/2006ApJ...648....7K} {648, 7}

\bibitem[\protect\citeauthoryear{{Kawabe}, {Kohno}, {Tamura}, {Takekoshi},
  {Oshima}  \& {Ishii}}{{Kawabe} et~al.}{2016}]{Kawabe16}
{Kawabe} R.,  {Kohno} K.,  {Tamura} Y.,  {Takekoshi} T.,  {Oshima} T.,
  {Ishii} S.,  2016, in Ground-based and Airborne Telescopes VI. p. 990626,
  \mn@doi{10.1117/12.2232202}

\bibitem[\protect\citeauthoryear{{Kennicutt}}{{Kennicutt}}{1998}]{Kennicutt98}
{Kennicutt} Robert~C. J.,  1998, \mn@doi [\araa]
  {10.1146/annurev.astro.36.1.189}, \href
  {https://ui.adsabs.harvard.edu/abs/1998ARA&A..36..189K} {36, 189}

\bibitem[\protect\citeauthoryear{{Knudsen}, {Richard}, {Kneib}, {Jauzac},
  {Cl{\'e}ment}, {Drouart}, {Egami}  \& {Lindroos}}{{Knudsen}
  et~al.}{2016}]{Knudsen16}
{Knudsen} K.~K.,  {Richard} J.,  {Kneib} J.-P.,  {Jauzac} M.,  {Cl{\'e}ment}
  B.,  {Drouart} G.,  {Egami} E.,   {Lindroos} L.,  2016, \mn@doi [\mnras]
  {10.1093/mnrasl/slw114}, \href
  {https://ui.adsabs.harvard.edu/abs/2016MNRAS.462L...6K} {462, L6}

\bibitem[\protect\citeauthoryear{{Kohandel}, {Ferrara}, {Pallottini},
  {Vallini}, {Sommovigo}  \& {Ziparo}}{{Kohandel} et~al.}{2023}]{Kohandel23}
{Kohandel} M.,  {Ferrara} A.,  {Pallottini} A.,  {Vallini} L.,  {Sommovigo} L.,
    {Ziparo} F.,  2023, \mn@doi [\mnras] {10.1093/mnrasl/slac166}, \href
  {https://ui.adsabs.harvard.edu/abs/2023MNRAS.520L..16K} {520, L16}

\bibitem[\protect\citeauthoryear{{Kovetz} et~al.,}{{Kovetz}
  et~al.}{2017}]{Kovetz17}
{Kovetz} E.~D.,  et~al., 2017, arXiv e-prints, \href
  {http://adsabs.harvard.edu/abs/2017arXiv170909066K} {p. arXiv:1709.09066}

\bibitem[\protect\citeauthoryear{{Kubota}, {Yoshiura}, {Takahashi}, {Hasegawa},
  {Yajima}, {Ouchi}, {Pindor}  \& {Webster}}{{Kubota} et~al.}{2018}]{Kubota18}
{Kubota} K.,  {Yoshiura} S.,  {Takahashi} K.,  {Hasegawa} K.,  {Yajima} H.,
  {Ouchi} M.,  {Pindor} B.,   {Webster} R.~L.,  2018, \mn@doi [\mnras]
  {10.1093/mnras/sty1471}, \href
  {http://adsabs.harvard.edu/abs/2018MNRAS.479.2754K} {479, 2754}

\bibitem[\protect\citeauthoryear{{La Plante}, {Mirocha}, {Gorce}, {Lidz}  \&
  {Parsons}}{{La Plante} et~al.}{2023}]{LaPlante23}
{La Plante} P.,  {Mirocha} J.,  {Gorce} A.,  {Lidz} A.,   {Parsons} A.,  2023,
  \mn@doi [\apj] {10.3847/1538-4357/acaeb0}, \href
  {https://ui.adsabs.harvard.edu/abs/2023ApJ...944...59L} {944, 59}

\bibitem[\protect\citeauthoryear{{Laporte} et~al.,}{{Laporte}
  et~al.}{2019}]{Laporte19}
{Laporte} N.,  et~al., 2019, \mn@doi [\mnras] {10.1093/mnrasl/slz094}, \href
  {https://ui.adsabs.harvard.edu/abs/2019MNRAS.487L..81L} {487, L81}

\bibitem[\protect\citeauthoryear{{Lidz}, {Zahn}, {McQuinn}, {Zaldarriaga},
  {Dutta}  \& {Hernquist}}{{Lidz} et~al.}{2007}]{Lidz07}
{Lidz} A.,  {Zahn} O.,  {McQuinn} M.,  {Zaldarriaga} M.,  {Dutta} S.,
  {Hernquist} L.,  2007, \mn@doi [\apj] {10.1086/511670}, \href
  {https://ui.adsabs.harvard.edu/abs/2007ApJ...659..865L} {659, 865}

\bibitem[\protect\citeauthoryear{{Lidz}, {Zahn}, {Furlanetto}, {McQuinn},
  {Hernquist}  \& {Zaldarriaga}}{{Lidz} et~al.}{2009}]{Lidz09}
{Lidz} A.,  {Zahn} O.,  {Furlanetto} S.~R.,  {McQuinn} M.,  {Hernquist} L.,
  {Zaldarriaga} M.,  2009, \mn@doi [\apj] {10.1088/0004-637X/690/1/252}, \href
  {http://adsabs.harvard.edu/abs/2009ApJ...690..252L} {690, 252}

\bibitem[\protect\citeauthoryear{{Lidz}, {Furlanetto}, {Oh}, {Aguirre},
  {Chang}, {Dor{\'e}}  \& {Pritchard}}{{Lidz} et~al.}{2011}]{Lidz11}
{Lidz} A.,  {Furlanetto} S.~R.,  {Oh} S.~P.,  {Aguirre} J.,  {Chang} T.-C.,
  {Dor{\'e}} O.,   {Pritchard} J.~R.,  2011, \mn@doi [\apj]
  {10.1088/0004-637X/741/2/70}, \href
  {https://ui.adsabs.harvard.edu/abs/2011ApJ...741...70L} {741, 70}

\bibitem[\protect\citeauthoryear{{Liu} \& {Shaw}}{{Liu} \&
  {Shaw}}{2020}]{Liu20}
{Liu} A.,  {Shaw} J.~R.,  2020, \mn@doi [\pasp] {10.1088/1538-3873/ab5bfd},
  \href {https://ui.adsabs.harvard.edu/abs/2020PASP..132f2001L} {132, 062001}

\bibitem[\protect\citeauthoryear{{Livermore}, {Finkelstein}  \&
  {Lotz}}{{Livermore} et~al.}{2017}]{Livermore17}
{Livermore} R.~C.,  {Finkelstein} S.~L.,   {Lotz} J.~M.,  2017, \mn@doi [\apj]
  {10.3847/1538-4357/835/2/113}, \href
  {https://ui.adsabs.harvard.edu/abs/2017ApJ...835..113L} {835, 113}

\bibitem[\protect\citeauthoryear{{Loeb} \& {Furlanetto}}{{Loeb} \&
  {Furlanetto}}{2013}]{Loeb13}
{Loeb} A.,  {Furlanetto} S.~R.,  2013, {The First Galaxies in the Universe}

\bibitem[\protect\citeauthoryear{{Madau} \& {Dickinson}}{{Madau} \&
  {Dickinson}}{2014}]{Madau14}
{Madau} P.,  {Dickinson} M.,  2014, \mn@doi [\araa]
  {10.1146/annurev-astro-081811-125615}, \href
  {https://ui.adsabs.harvard.edu/abs/2014ARA&A..52..415M} {52, 415}

\bibitem[\protect\citeauthoryear{{Masui} et~al.,}{{Masui}
  et~al.}{2013}]{Masui13}
{Masui} K.~W.,  et~al., 2013, \mn@doi [\apjl] {10.1088/2041-8205/763/1/L20},
  \href {https://ui.adsabs.harvard.edu/abs/2013ApJ...763L..20M} {763, L20}

\bibitem[\protect\citeauthoryear{{Mertens} et~al.,}{{Mertens}
  et~al.}{2020}]{Mertens20}
{Mertens} F.~G.,  et~al., 2020, \mn@doi [\mnras] {10.1093/mnras/staa327}, \href
  {https://ui.adsabs.harvard.edu/abs/2020MNRAS.493.1662M} {493, 1662}

\bibitem[\protect\citeauthoryear{{Mesinger} \& {Furlanetto}}{{Mesinger} \&
  {Furlanetto}}{2007}]{Mesinger07}
{Mesinger} A.,  {Furlanetto} S.,  2007, \mn@doi [\apj] {10.1086/521806}, \href
  {https://ui.adsabs.harvard.edu/abs/2007ApJ...669..663M} {669, 663}

\bibitem[\protect\citeauthoryear{{Mesinger}, {Furlanetto}  \& {Cen}}{{Mesinger}
  et~al.}{2011}]{Mesinger11}
{Mesinger} A.,  {Furlanetto} S.,   {Cen} R.,  2011, \mn@doi [\mnras]
  {10.1111/j.1365-2966.2010.17731.x}, \href
  {https://ui.adsabs.harvard.edu/abs/2011MNRAS.411..955M} {411, 955}

\bibitem[\protect\citeauthoryear{{Moriwaki} et~al.,}{{Moriwaki}
  et~al.}{2018}]{Moriwaki18}
{Moriwaki} K.,  et~al., 2018, \mn@doi [\mnras] {10.1093/mnrasl/sly167}, \href
  {http://adsabs.harvard.edu/abs/2018MNRAS.481L..84M} {481, L84}

\bibitem[\protect\citeauthoryear{{Moriwaki}, {Yoshida}, {Eide}  \&
  {Ciardi}}{{Moriwaki} et~al.}{2019}]{Moriwaki19}
{Moriwaki} K.,  {Yoshida} N.,  {Eide} M.~B.,   {Ciardi} B.,  2019, \mn@doi
  [\mnras] {10.1093/mnras/stz2308}, \href
  {https://ui.adsabs.harvard.edu/abs/2019MNRAS.489.2471M} {489, 2471}

\bibitem[\protect\citeauthoryear{{Murmu}, {Majumdar}  \& {Datta}}{{Murmu}
  et~al.}{2021}]{Murmu21}
{Murmu} C.~S.,  {Majumdar} S.,   {Datta} K.~K.,  2021, \mn@doi [\mnras]
  {10.1093/mnras/stab2347}, \href
  {https://ui.adsabs.harvard.edu/abs/2021MNRAS.507.2500M} {507, 2500}

\bibitem[\protect\citeauthoryear{{Murmu} et~al.,}{{Murmu}
  et~al.}{2023}]{Murmu23}
{Murmu} C.~S.,  et~al., 2023, \mn@doi [\mnras] {10.1093/mnras/stac3304}, \href
  {https://ui.adsabs.harvard.edu/abs/2023MNRAS.518.3074M} {518, 3074}

\bibitem[\protect\citeauthoryear{Murray, Greig, Mesinger, Muñoz, Qin, Park  \&
  Watkinson}{Murray et~al.}{2020}]{Murray20}
Murray S.~G.,  Greig B.,  Mesinger A.,  Muñoz J.~B.,  Qin Y.,  Park J.,
  Watkinson C.~A.,  2020, \mn@doi [Journal of Open Source Software]
  {10.21105/joss.02582}, 5, 2582

\bibitem[\protect\citeauthoryear{{Nakajima}, {Ouchi}, {Shimasaku}, {Hashimoto},
  {Ono}  \& {Lee}}{{Nakajima} et~al.}{2013}]{Nakajima13}
{Nakajima} K.,  {Ouchi} M.,  {Shimasaku} K.,  {Hashimoto} T.,  {Ono} Y.,
  {Lee} J.~C.,  2013, \mn@doi [\apj] {10.1088/0004-637X/769/1/3}, \href
  {https://ui.adsabs.harvard.edu/abs/2013ApJ...769....3N} {769, 3}

\bibitem[\protect\citeauthoryear{{Nakazato}, {Yoshida}  \&
  {Ceverino}}{{Nakazato} et~al.}{2023}]{Nakazato23}
{Nakazato} Y.,  {Yoshida} N.,   {Ceverino} D.,  2023, \mn@doi [\apj]
  {10.3847/1538-4357/ace25a}, \href
  {https://ui.adsabs.harvard.edu/abs/2023ApJ...953..140N} {953, 140}

\bibitem[\protect\citeauthoryear{Neben, Stalder, Hewitt  \& Tonry}{Neben
  et~al.}{2017}]{Neben17}
Neben A.~R.,  Stalder B.,  Hewitt J.~N.,   Tonry J.~L.,  2017, \mn@doi [The
  Astrophysical Journal] {10.3847/1538-4357/aa8f9c}, 849, 50

\bibitem[\protect\citeauthoryear{{Nelson} et~al.,}{{Nelson}
  et~al.}{2019}]{Nelson19}
{Nelson} D.,  et~al., 2019, \mn@doi [Computational Astrophysics and Cosmology]
  {10.1186/s40668-019-0028-x}, \href
  {https://ui.adsabs.harvard.edu/abs/2019ComAC...6....2N} {6, 2}

\bibitem[\protect\citeauthoryear{{Paciga} et~al.,}{{Paciga}
  et~al.}{2013}]{Paciga13}
{Paciga} G.,  et~al., 2013, \mn@doi [\mnras] {10.1093/mnras/stt753}, \href
  {https://ui.adsabs.harvard.edu/abs/2013MNRAS.433..639P} {433, 639}

\bibitem[\protect\citeauthoryear{{Padmanabhan}}{{Padmanabhan}}{2023}]{Padmanabhan23}
{Padmanabhan} H.,  2023, \mn@doi [\mnras] {10.1093/mnras/stad1559}, \href
  {https://ui.adsabs.harvard.edu/abs/2023MNRAS.523.3503P} {523, 3503}

\bibitem[\protect\citeauthoryear{{Park}, {Kim}, {Wyithe}  \& {Lacey}}{{Park}
  et~al.}{2014}]{Park14}
{Park} J.,  {Kim} H.-S.,  {Wyithe} J. S.~B.,   {Lacey} C.~G.,  2014, \mn@doi
  [\mnras] {10.1093/mnras/stt2366}, \href
  {https://ui.adsabs.harvard.edu/abs/2014MNRAS.438.2474P} {438, 2474}

\bibitem[\protect\citeauthoryear{{Park}, {Mesinger}, {Greig}  \&
  {Gillet}}{{Park} et~al.}{2019}]{Park19}
{Park} J.,  {Mesinger} A.,  {Greig} B.,   {Gillet} N.,  2019, \mn@doi [\mnras]
  {10.1093/mnras/stz032}, \href
  {https://ui.adsabs.harvard.edu/abs/2019MNRAS.484..933P} {484, 933}

\bibitem[\protect\citeauthoryear{{Patil} et~al.,}{{Patil}
  et~al.}{2017}]{Patil17}
{Patil} A.~H.,  et~al., 2017, \mn@doi [\apj] {10.3847/1538-4357/aa63e7}, \href
  {https://ui.adsabs.harvard.edu/abs/2017ApJ...838...65P} {838, 65}

\bibitem[\protect\citeauthoryear{{P{\'e}rez-Gonz{\'a}lez}
  et~al.,}{{P{\'e}rez-Gonz{\'a}lez} et~al.}{2023}]{Perez-Gonzalez23}
{P{\'e}rez-Gonz{\'a}lez} P.~G.,  et~al., 2023, \mn@doi [\apjl]
  {10.3847/2041-8213/acd9d0}, \href
  {https://ui.adsabs.harvard.edu/abs/2023ApJ...951L...1P} {951, L1}

\bibitem[\protect\citeauthoryear{{Planck Collaboration VI}}{{Planck
  Collaboration VI}}{2018}]{Planck18}
{Planck Collaboration VI} 2018, arXiv e-prints, \href
  {https://ui.adsabs.harvard.edu/abs/2018arXiv180706209P} {p. arXiv:1807.06209}

\bibitem[\protect\citeauthoryear{{Pritchard} \& {Furlanetto}}{{Pritchard} \&
  {Furlanetto}}{2007}]{Pritchard07}
{Pritchard} J.~R.,  {Furlanetto} S.~R.,  2007, \mn@doi [\mnras]
  {10.1111/j.1365-2966.2007.11519.x}, \href
  {https://ui.adsabs.harvard.edu/abs/2007MNRAS.376.1680P} {376, 1680}

\bibitem[\protect\citeauthoryear{{Pritchard} \& {Loeb}}{{Pritchard} \&
  {Loeb}}{2008}]{Pritchard08}
{Pritchard} J.~R.,  {Loeb} A.,  2008, \mn@doi [\prd]
  {10.1103/PhysRevD.78.103511}, \href
  {https://ui.adsabs.harvard.edu/abs/2008PhRvD..78j3511P} {78, 103511}

\bibitem[\protect\citeauthoryear{{Pritchard} \& {Loeb}}{{Pritchard} \&
  {Loeb}}{2012}]{Pritchard12}
{Pritchard} J.~R.,  {Loeb} A.,  2012, \mn@doi [Reports on Progress in Physics]
  {10.1088/0034-4885/75/8/086901}, \href
  {https://ui.adsabs.harvard.edu/abs/2012RPPh...75h6901P} {75, 086901}

\bibitem[\protect\citeauthoryear{{Reis}, {Barkana}  \& {Fialkov}}{{Reis}
  et~al.}{2022}]{Reis22}
{Reis} I.,  {Barkana} R.,   {Fialkov} A.,  2022, \mn@doi [\mnras]
  {10.1093/mnras/stac411}, \href
  {https://ui.adsabs.harvard.edu/abs/2022MNRAS.511.5265R} {511, 5265}

\bibitem[\protect\citeauthoryear{{Schaerer} et~al.,}{{Schaerer}
  et~al.}{2020}]{Schaerer20}
{Schaerer} D.,  et~al., 2020, \mn@doi [\aap] {10.1051/0004-6361/202037617},
  \href {https://ui.adsabs.harvard.edu/abs/2020A&A...643A...3S} {643, A3}

\bibitem[\protect\citeauthoryear{{Seljak}, {Hamaus}  \& {Desjacques}}{{Seljak}
  et~al.}{2009}]{Seljak09}
{Seljak} U.,  {Hamaus} N.,   {Desjacques} V.,  2009, \mn@doi [\prl]
  {10.1103/PhysRevLett.103.091303}, \href
  {https://ui.adsabs.harvard.edu/abs/2009PhRvL.103i1303S} {103, 091303}

\bibitem[\protect\citeauthoryear{{Shimabukuro}, {Yoshiura}, {Takahashi},
  {Yokoyama}  \& {Ichiki}}{{Shimabukuro} et~al.}{2017}]{Shimabukuro17}
{Shimabukuro} H.,  {Yoshiura} S.,  {Takahashi} K.,  {Yokoyama} S.,   {Ichiki}
  K.,  2017, \mn@doi [\mnras] {10.1093/mnras/stx530}, \href
  {https://ui.adsabs.harvard.edu/abs/2017MNRAS.468.1542S} {468, 1542}

\bibitem[\protect\citeauthoryear{{Silva}, {Santos}, {Gong}, {Cooray}  \&
  {Bock}}{{Silva} et~al.}{2013}]{Silva13}
{Silva} M.~B.,  {Santos} M.~G.,  {Gong} Y.,  {Cooray} A.,   {Bock} J.,  2013,
  \mn@doi [\apj] {10.1088/0004-637X/763/2/132}, \href
  {https://ui.adsabs.harvard.edu/abs/2013ApJ...763..132S} {763, 132}

\bibitem[\protect\citeauthoryear{{Spergel} et~al.,}{{Spergel}
  et~al.}{2015}]{Spergel15}
{Spergel} D.,  et~al., 2015, \mn@doi [arXiv e-prints]
  {10.48550/arXiv.1503.03757}, \href
  {https://ui.adsabs.harvard.edu/abs/2015arXiv150303757S} {p. arXiv:1503.03757}

\bibitem[\protect\citeauthoryear{{Tanaka} \& {Hasegawa}}{{Tanaka} \&
  {Hasegawa}}{2021}]{Tanaka21}
{Tanaka} T.,  {Hasegawa} K.,  2021, \mn@doi [\mnras] {10.1093/mnras/stab072},
  \href {https://ui.adsabs.harvard.edu/abs/2021MNRAS.502..463T} {502, 463}

\bibitem[\protect\citeauthoryear{{Trott} et~al.,}{{Trott}
  et~al.}{2020}]{Trott20}
{Trott} C.~M.,  et~al., 2020, \mn@doi [\mnras] {10.1093/mnras/staa414}, \href
  {https://ui.adsabs.harvard.edu/abs/2020MNRAS.493.4711T} {493, 4711}

\bibitem[\protect\citeauthoryear{{Vrbanec} et~al.,}{{Vrbanec}
  et~al.}{2016}]{Vrbanec16}
{Vrbanec} D.,  et~al., 2016, \mn@doi [\mnras] {10.1093/mnras/stv2993}, \href
  {http://adsabs.harvard.edu/abs/2016MNRAS.457..666V} {457, 666}

\bibitem[\protect\citeauthoryear{{Wang} et~al.,}{{Wang} et~al.}{2022}]{Wang22}
{Wang} Y.,  et~al., 2022, \mn@doi [\apj] {10.3847/1538-4357/ac4973}, \href
  {https://ui.adsabs.harvard.edu/abs/2022ApJ...928....1W} {928, 1}

\bibitem[\protect\citeauthoryear{{Weinberger}, {Kulkarni}  \&
  {Haehnelt}}{{Weinberger} et~al.}{2020}]{Weinberger20}
{Weinberger} L.~H.,  {Kulkarni} G.,   {Haehnelt} M.~G.,  2020, \mn@doi [\mnras]
  {10.1093/mnras/staa749}, \href
  {https://ui.adsabs.harvard.edu/abs/2020MNRAS.494..703W} {494, 703}

\bibitem[\protect\citeauthoryear{{Wiersma} et~al.,}{{Wiersma}
  et~al.}{2013}]{Wiersma13}
{Wiersma} R.~P.~C.,  et~al., 2013, \mn@doi [\mnras] {10.1093/mnras/stt624},
  \href {http://adsabs.harvard.edu/abs/2013MNRAS.432.2615W} {432, 2615}

\bibitem[\protect\citeauthoryear{{Wolz} et~al.,}{{Wolz} et~al.}{2022}]{Wolz22}
{Wolz} L.,  et~al., 2022, \mn@doi [\mnras] {10.1093/mnras/stab3621}, \href
  {https://ui.adsabs.harvard.edu/abs/2022MNRAS.510.3495W} {510, 3495}

\bibitem[\protect\citeauthoryear{{Wyithe} \& {Morales}}{{Wyithe} \&
  {Morales}}{2007}]{Wyithe07}
{Wyithe} J. S.~B.,  {Morales} M.~F.,  2007, \mn@doi [\mnras]
  {10.1111/j.1365-2966.2007.12048.x}, \href
  {https://ui.adsabs.harvard.edu/abs/2007MNRAS.379.1647W} {379, 1647}

\bibitem[\protect\citeauthoryear{Yoshiura, Line, Kubota, Hasegawa  \&
  Takahashi}{Yoshiura et~al.}{2018}]{Yoshiura18}
Yoshiura S.,  Line J. L.~B.,  Kubota K.,  Hasegawa K.,   Takahashi K.,  2018,
  \mn@doi [Monthly Notices of the Royal Astronomical Society]
  {10.1093/mnras/sty1472}, 479, 2767

\bibitem[\protect\citeauthoryear{{Yoshiura} et~al.,}{{Yoshiura}
  et~al.}{2021}]{Yoshiura21}
{Yoshiura} S.,  et~al., 2021, \mn@doi [\mnras] {10.1093/mnras/stab1560}, \href
  {https://ui.adsabs.harvard.edu/abs/2021MNRAS.505.4775Y} {505, 4775}

\makeatother
\end{thebibliography}

%\begin{thebibliography}{99}
%\end{thebibliography}

%%%%%%%%%%%%%%%%%%%%%%%%%%%%%%%%%%%%%%%%%%%%%%%%%%

%%%%%%%%%%%%%%%%% APPENDICES %%%%%%%%%%%%%%%%%%%%%

\appendix

%%%%%%%%%%%%%%%%%%%%%%%%%%%%%%%%%%%%%%%%%%%%%%%%%%

% Don't change these lines
\bsp	% typesetting comment
\label{lastpage}
\end{document}